\pdfoutput=1
\documentclass[twocolumn]{aastex62}

\newcommand\aastex{AAS\TeX}
\newcommand\latex{La\TeX}

\newcommand{\ms}[1]{\textcolor{black}{#1}}

\newcommand{\beq}{\begin{equation}}
\newcommand{\beqa}{\begin{eqnarray}}
\newcommand{\eeq}{\end{equation}}
\newcommand{\eeqa}{\end{eqnarray}}

\newcommand{\simgt}{\lower.5ex\hbox{$\; \buildrel > \over \sim \;$}}
\newcommand{\simlt}{\lower.5ex\hbox{$\; \buildrel < \over \sim \;$}}

\newcommand{\bd}[1]{\mbox{\boldmath $#1$}}

\graphicspath{{./}{figures/}}

\submitjournal{ApJ}

\shorttitle{Pseudo evolution of galaxy-cluster masses}
\shortauthors{Shirasaki}
\begin{document}

\title{\ms{The} Pseudo Evolution of Galaxy-Cluster Masses and Its \ms{Connection} to Mass Density Profile}

\correspondingauthor{Masato Shirasaki}
\email{masato.shirasaki@nao.ac.jp}

%\author[0000-0002-0786-7307]{Greg J. Schwarz}
%\affil{American Astronomical Society \\
%2000 Florida Ave., NW, Suite 300 \\
%Washington, DC 20009-1231, USA}

\author{Masato Shirasaki}
\affiliation{National Astronomical Observatory of Japan, Mitaka, Tokyo 181-8588, Japan}



\begin{abstract}

%250 word limit for the abstract
A mass of \ms{dark matter halo} is commonly defined as the spherical over-density (SO) mass with respect to a reference density, whereas the time evolution of an SO mass can be affected by the redshift evolution of the reference density
as well as the physical mass accretion around \ms{halos}. 
In this study, we directly measure the amount of pseudo evolution of the SO masses of cluster-sized halos 
by the changes in the reference density from a time series of $N$-body simulations for the first time. 
We find that the \ms{$52\pm19\%$} difference in the virial SO masses between $z=0$ and $1$ 
can be accounted for by the pseudo evolution of clusters with a virial mass of $10^{14}\, h^{-1}M_{\odot}$ at $z=0$. 
The amount of pseudo evolution is found to be correlated with the age and density environment
of a galaxy cluster. The stacked mass density profiles of cluster-sized halos 
with a greater amount of pseudo evolution in the SO mass shows the higher concentration ${\it and}$ greater linear bias parameter 
that is a counter-example of the known secondary halo bias due to concentration on the scale of clusters. 
We discuss how more concentrated clusters can show larger clustering amplitudes than their less concentrated counterparts and argue that the presence of rich filamentary structures plays a critical role in determining 
the linear halo bias of galaxy clusters.
\end{abstract}



\section{Introduction} \label{sec:intro}

% galaxy cluster, abundance for cosmology
Galaxy clusters are known to be the largest self-bound objects in the universe,
and the statistical properties of these clusters are of significant importance in modern cosmology.
The number density of galaxy clusters as a function of mass
is expected to be sensitive to the gravitational growth in linear density fluctuations and the average cosmic mass density \citep[e.g.][]{2011ARA&A..49..409A}.
Hence, observations on the abundance of galaxy clusters 
allow the establishment of the standard cosmological model \citep[e.g.][]{2009ApJ...692.1060V, 2010ApJ...708..645R, 2014MNRAS.440.2077M, 2016A&A...594A..24P, 2016ApJ...832...95D}, 
but current cosmological constraints hinge 
on the accuracy in the estimation of mass.

% need to estimate the cluster mass, accurately.
% the stacked lensing analysis is commonly adopted to determine the average mass of a sample of clusters
The cross-correlation of galaxy clusters with 
the shapes of background galaxies, referred to as stacked lensing, 
is a unique means of measuring the average total matter distribution 
of the clusters in the foreground.
Stacked lensing analysis has the significant advantage of being 
independent from the dynamical state of a galaxy cluster, 
enabling the direct measurement of the gravitating mass of a cluster.
It has been already applied to observational data sets
and provided meaningful constraints on the relationship 
between multi-wavelength observables and the underlying cluster masses
\citep[e.g][]{2007arXiv0709.1159J, 2011ApJ...738...41U, 2013ApJ...769L..35O, 2017MNRAS.466.3103S, 2018PASJ...70S..28M, 2018ApJ...854..120M, 2019ApJ...875...63M}.
In addition, large-scale amplitudes seen within the stacked lensing signal 
are expected to be associated with the clustering of dark matter halos of similar size \citep[e.g.][]{2014ApJ...784L..25C}.
A combined analysis of the cluster abundance with the stacked lensing signal including the scales beyond
a virial regime can be a powerful probe of the physics that lies beyond that of the standard model, such as 
massive neutrinos, the origin of the accelerating expansion of the universe,
and the inflationary physics of the early universe \citep[e.g.][]{OguriTakada:11, 2016PASJ...68....4S, 2017PhRvD..96j3525M}.

% linear bias would be affected by selection method of galaxy clusters -> assembly bias
The mass density profile around a galaxy cluster is relevant to the stacked lensing analysis
and accurate modeling of this profile is therefore essential for future cosmological analysis using galaxy clusters.
It is commonly assumed that the density profile of a cluster solely depends on the mass 
at a given redshift, but this is not always valid.
Numerical studies have suggested that the density profiles of dark matter halos can be expressed 
in a single universal form \citep[e.g.][]{Navarroetal:97}, but the characteristic scales in the universal profiles will depend 
on the history of the mass accretion of the halos \citep[e.g.][]{2001MNRAS.321..559B, 2002ApJ...568...52W, 2013MNRAS.432.1103L}.
In addition, the large-scale amplitudes found within mass density profiles are usually parametrized 
with a linear bias parameter and the halo biases of clusters can depend not only on their masses, 
but also on other properties defined in their inner density profiles \citep[e.g.][]{2007MNRAS.377L...5G, 2010ApJ...708..469F}.
On the scale of galaxy-clusters, 
the extent to which the mass assembly history can change the halo bias is still controversial.
The latest cosmological simulations enable us to study the dependence of secondary parameters (other than mass) on the halo bias and it is apparent that the clustering of large halos is insensitive 
to the half-mass redshift \citep{2018MNRAS.474.5143M}, but a detailed characterization of the mass assembly history may be required to assess the assembly bias of cluster-sized halos \citep{2018JCAP...10..012C}.

% spherical over-density (SO) mass is commonly adopted in the study of mass assembly history, but it would be affected by redshift evolution of reference density
% No previous studies have measured the amount of pseudo evolution in SO mass -> we did for the first time
As pointed out in the literature \citep[e.g.][]{2014ApJ...792..124Z, 2018JCAP...10..012C}, 
one issue in studying mass assembly history is the definition of halo mass in numerical simulations.
The spherical over-density (SO) mass has commonly been adopted in previous studies 
of mass assembly history, defined by
\beqa
M_{\Delta}(z) = \frac{4}{3}\pi \, R^3_{\Delta}(z)\, \Delta(z) \rho_{\rm ref}(z), \label{eq:def_SO_mass}
\eeqa
where $\Delta(z)$ is an SO parameter and $\rho_{\rm ref}(z)$ is a reference density.
\citet{2013ApJ...766...25D} have derived that the time evolution of an SO mass can 
depend on the physical mass accretion outside the dark matter halos and on 
the evolution in the redshift of the reference density. 
Even if a dark matter halo has a completely static density profile with regards to its physical coordinates 
and there are no physical mass accretions,
the SO mass can evolve over time. This evolution is referred to as the pseudo evolution.
%To the best of our knowledge, no previous numerical studies have measured the amount of pseudo evolution of 
%SO mass of cluster-sized halo so far \citep[but, see][for the robust constraint of the amount of pseudo evolution on cluster scales]{2013ApJ...766...25D}.
\ms{There have been numerous efforts to quantify and describe pseudo evolution using numerical simulations in the literature.}
\ms{\citet{2008MNRAS.389..385C} introduced a new halo boundary defined by the innermost radius at which there is zero mean radial velocity and then studied the relationship between the new boundaries and the commonly adopted SO radii.}
\ms{\citet{2013ApJ...766...25D} have constructed a novel means of estimating the amount of pseudo evolution
using the halo density profiles at two different redshifts.}
\ms{\citet{2014ApJ...792..124Z} has investigated the pseudo evolution for various SO mass definitions 
by using an enclosed mass with a fixed physical scale.}
\ms{\citet{2015ApJ...808...40W} have explored the redshift evolution of the density profiles of dark matter 
and baryons in the context of pseudo evolution, whereas \citet{2016ApJ...818..188D} have derived a relationship between the mass profile and mass accretion rate, and validated this relationship with numerical simulations.}

It is important to study the amount of pseudo evolution that has occurred in the SO masses to refine our understanding
of the mass assembly history of massive dark matter halos, 
and to develop a more accurate model of the stacked lensing signals from galaxy clusters.
\ms{In particular, most previous studies have relied on numerical simulations with a small volume 
and/or a large time interval, leading to a situation where a detailed description of the pseudo evolution of individual cluster-sized halos is still unclear.}
In this study, we measure the amount of pseudo evolution that has taken place to the SO masses of 
cluster-sized halos by tracking the time evolution of their density profiles along with their merger histories 
in $N$-body simulations with a fine snapshot spacing.
We also study the correlation among the pseudo evolution, the concentration of the inner density profile, 
age, and the density of the environment surrounding the halos.
We then quantify the relationship between the amount of pseudo evolution and the averaged mass density profile, 
which is essential for the practical application of stacked lensing analysis for galaxy clusters.

The paper is organized as follows. In Section~\ref{sec:Basics},
an overview of the time evolution of the SO mass and introduce two common definitions for SO mass.
In Section~\ref{sec:sims}, we describe the $N$-body simulations and how to quantify the amount of pseudo evolution in the simulations. The results are presented in Section~\ref{sec:res}. Finally, the 
conclusions and discussions are provided in Section~\ref{sec:concl}.

\section{Time evolution of enclosed mass within a sphere} \label{sec:Basics}

To study the mass distribution around galaxy clusters, 
a spherically symmetric mass density profile is considered $\rho(r, z)$, 
where $r$ is the cluster-centric radius and $z$ is the redshift under investigation.
Using this spherical density profile, the SO mass can be defined (Eq.~[\ref{eq:def_SO_mass}]) as
\beqa
M_{\Delta}(z) = \int_{0}^{R_{\Delta}(z)} 4\pi r^2 {\rm d}r \, \rho(r, z).
\eeqa

As introduced in \citet{2013ApJ...766...25D}, the redshift evolution of an SO mass $M_{\Delta}$ can be
decomposed into two terms:
\beqa
\frac{{\rm d}M_{\Delta}(z)}{{\rm d}z} &=& 
4 \pi R^2_{\Delta}(z)\rho(R_{\Delta},z)\frac{{\rm d}R_{\Delta}(z)}{{\rm d}z}  \nonumber \\
&&\qquad \qquad  + \int_{0}^{R_{\Delta}(z)} 4\pi r^2 {\rm d}r \, \frac{{\rm d} \rho(r, z)}{{\rm d}z}, \label{eq:evol_SO_mass}
\eeqa
where the first term in the right-hand side represents the pseudo evolution caused by the change in SO radius 
and the second term determines the evolution of the mass due to the actual growth of the density profile. 
The amount of pseudo evolution in the actual redshift evolution of an SO mass would be 
an important integrand in understanding the mass accretion history of a dark matter halo on a physical basis. 
\citet{2013ApJ...766...25D} have studied the amount of pseudo evolution between $z=0$ and $1$
for a wide range of halo masses with $N$-body simulations. It would be worth noting that
the results in \citet{2013ApJ...766...25D} are still based on the density profiles of halos 
at $z=0$ and their progenitors at $z=1$. Although \citet{2013ApJ...766...25D} 
included lower and upper limits, a direct estimate of the amount of pseudo evolution that has occurred on the scale of galaxy-cluster 
is yet to be obtained \citep[but, see][for the direct estimate on galaxy scales]{2007ApJ...667..859D}. 
In this study, we measure the amount of pseudo evolution that has taken place in an SO mass between two redshifts
by using $\sim30$ snapshots of $N$-body simulations.

There are two conventional choices for the reference density in Eq.~(\ref{eq:def_SO_mass}).
One is the mean matter density which is denoted as $\bar{\rho}_{\rm m}$ in this study, and another is 
the critical density of the universe denoted as $\rho_{\rm c}$. Throughout this study, 
we use the definition of SO mass with respect to 200 times $\bar{\rho}_{\rm m}$
as well as the virial SO mass which is represented by $\Delta_{\rm vir}(z)$ times $\rho_{\rm c}$,
where $\Delta_{\rm vir}$ is given by the spherical collapse model \citep{1998ApJ...495...80B}.
For the former and the latter, we denote $M_{\rm 200b}$ and $M_{\rm vir}$, respectively.

\section{Method} \label{sec:sims}

\subsection{Simulation design} \label{subsec:design}

We run cosmological $N$-body simulations to study the time evolution of the density profiles 
around the galaxy clusters.
We use the parallel Tree-Particle Mesh code {\tt GADGET2} \citep{2005MNRAS.364.1105S}.
With $1024^3$ dark matter particles per unit volume of $500\, h^{-1}\, {\rm Mpc}$ on one side. 
We generate the initial conditions using a parallel code developed by \citet{2009PASJ...61..321N} and \citet{2011A&A...527A..87V}, which employs the second-order Lagrangian perturbation theory \citep[e.g.][]{2006MNRAS.373..369C}. The initial redshift is set to $z_{\rm init}=59$, where we compute the linear matter transfer function using {\tt CAMB} \citep{Lewis:1999bs}. 
We adopt the following parameters for the simulations: 
present-day matter density parameter $\Omega_{\rm m0} = 0.3156$, 
dark energy density $\Omega_{\Lambda} = 1-\Omega_{\rm m0} = 0.6844$, 
the density fluctuation amplitude $\sigma_{8} = 0.831$, 
the parameter of the equation for the state of dark energy $w_{0}=-1$, 
Hubble parameter $h=0.6727$, 
and the scalar spectral index $n_s=0.9645$. 
These parameters are consistent with the results from Planck 2015 \citep{2016A&A...594A..13P}. 
In our simulation, the particle mass is set at $1.020\times10^{10}\, h^{-1}M_{\odot}$, 
with a softening length of $20.4\, h^{-1}{\rm kpc}$, which corresponds to 5 percent of the mean length of separation between particles.
Note that the simulation setup is the same as that used for the fiducial high-resolution run 
in \citet{2018arXiv181109504N}, except for the number of particles and the box length on one side.
The impact from the difference in generating initial conditions and mass resolution on halo statistics 
has been studied in \citet{2018arXiv181109504N} in detail. 
Simulation parameters such as the time-integral accuracy have been calibrated 
to enable the resulting matter power spectrum to converge within a sub-percent level \citep{2012ApJ...761..152T}.

The primary purpose of this study is to reveal the extent to which pseudo evolution 
can account for the actual redshift evolution of an SO mass between a given range of redshifts. 
As shown in Eq.~(\ref{eq:evol_SO_mass}), we require a fine redshift resolution 
for the density profile $\rho(r,z)$ to accurately measure the pseudo evolution of an SO mass.
\ms{To set the redshift interval between neighboring snapshots, we follow
the method in \citet{2017ApJS..231....5D}. 
%The method sets the snapshot spacing 
%in unit of the dynamical time of dark matter halos, which is defined by
%to range 
%from $6-15$ percent of the dynamical time of dark matter halos, which is defined by
We find the fiducial snapshot spacing in \citet{2017ApJS..231....5D} (i.e. the blue line in Figure 1 in \citet{2017ApJS..231....5D}), which can be approximated as
\beqa
t_{\rm dyn}(z) / \Delta t_{\rm snap} (z) &=& 
9.03\times10^{-4} \left[t_{\rm H}(z)/{\rm Gyr}\right]^{3} \nonumber \\
&&+1.4\times 10^{-2} \left[t_{\rm H}(z)/{\rm Gyr}\right]^{2} \nonumber \\
&&+6.0\times 10^{-2} \left[t_{\rm H}(z)/{\rm Gyr}\right] \nonumber \\
&& \qquad \qquad \qquad \qquad +6.55, \label{eq:dt_snapshot}
\eeqa
where $t_{\rm H}(z)=H^{-1}(z)$, $\Delta t_{\rm snap}$ is the snapshot spacing, and $t_{\rm dyn}$ is the dynamical time over which dark matter halos are present, which is defined by
\beqa
t_{\rm dyn}(z) = \frac{2R_{\Delta}}{v_{\Delta}} = 2^{3/2} H^{-1}(z) \left(\frac{\rho_{\Delta\, {\rm ref}}(z)}{\rho_{c}(z)}\right)^{-1/2},
\eeqa
where $v_{\Delta}$ represents the circular velocity at a radius of $R_{\Delta}$.
We use Eq.~(\ref{eq:dt_snapshot}) to set the snapshot spacing within the unit for dynamical time 
with $\rho_{\Delta\, {\rm ref}} = 200\bar{\rho}_{\rm m}$.}
When setting the initial redshift to 20, 
we find that a total of 110 snapshots are available until $z=0$.
\ms{Our snapshot spacing roughly corresponds to $10\%$ of the dynamical time in the redshift range of $0<z<1$.}
Note that the above method concerning the spacing of snapshots has been calibrated to provide sufficient information to determine the location of the first apocenter of the $N$-body particles after the infall 
into a halo, or the splashback \citep[see,][for details]{2017ApJS..231....5D}. 
Previous studies have shown that the splashback radius is in the range of $0.8-1.6$ 
times $R_{\rm 200b}$, but this also depends on the mass accretion rate \citep[e.g.,][]{2015ApJ...810...36M}. 
This indicates that our snapshot spacing will be adequate to track the time evolution of the mass density distribution in the virial region for individual halos \ms{(but see Appendix~\ref{app:sys_err_f_pseudo} for possible systematic uncertainties on the estimate of pseudo evolution in this study)}.

\subsection{Halo catalogs and merger tree} \label{subsec:catalog}

We use the phase-space halo finder {\tt ROCKSTAR} \citep{2013ApJ...762..109B} 
to identify dark matter halos in the simulation
and run the {\tt CONSISTENT-TREES} code \citep{2013ApJ...763...18B} 
to generate the merger trees from the {\tt ROCKSTAR} outputs.
In the following analyses, we consider the time evolution of the density profiles of isolated halos at 
$z=0$ with a peak height $\nu$ of greater than 1.5, 
except for the analysis in Section~\ref{subsec:diff_m_z}\footnote{In Section~\ref{subsec:diff_m_z}, we consider the amount of pseudo evolution for halos at the different redshifts of $z=0.3$, $0.6$ and $1$.}.
The peak height is commonly used to study the statistical properties of dark matter halos 
\citep[e.g.][]{1974ApJ...187..425P} and is defined by $\nu = \delta_{\rm c}/\sigma(M, z)$,
where $\delta_{\rm c}=1.686$ is the critical overdensity in the spherical top hat collapse model \citep{1972ApJ...176....1G},
and $\sigma(M,z)$ is the variance of smoothed linear density fluctuations at a specific redshift $z$
by a spherical top-hat filter. The top-hat radius $R_{\rm TH}$ in computing $\sigma(M, z)$ is set to 
$M_{\rm 200b}=4\pi/3 \times \bar{\rho}_{\rm m}(z=0) R_{\rm TH}^{3}$.
%\ms{In the following, we omit the argument of $z$ in $\sigma(M, z)$ for $z=0$ unless stated.}
The use of $M_{\rm 200b}$ to define the peak height is motivated by 
recent calibrations of halo statistics with a set of numerical simulations \citep[e.g.][]{2008ApJ...688..709T}
and \ms{the universality of outer density profiles \citep{2014ApJ...789....1D}.}
%and observational constraints of cluster mass with weak lensing analysis \citep[e.g.][]{2018ApJ...854..120M}.
In our adopted cosmological model, the peak height of $\nu=1.5$ corresponds to 
$M_{\rm 200b} = 4.6 \times10^{13}\, h^{-1}M_{\odot}$, which is resolved over 4600 particles in our simulation.
For the progenitor of a distinct halo, we define the halo along its most massive progenitor branch at each redshift.
We then construct the mass assembly history for each halo at $z=0$ by 
using the virial mass $M_{\rm vir}$ of its progenitor.

\subsection{Quantifying the amount of pseudo evolution} \label{subsec:stats}

To study the pseudo evolution of an SO mass, we extract the spherically averaged density profiles $\rho(r)$ and the enclosed spherical mass $M(<r)$ of halos 
in 80 logarithmically spaced bins between $0.05 R_{\rm 200b}$ and $10\, R_{\rm 200b}$. 
%We also measure $\rho(r)$ and $M(<r)$ of the progenitors in the same way.
Note that we include all the particles around halos within a sphere of $10\, R_{\rm 200b}$
when measuring $\rho(r)$ and $M(<r)$.
For a given halo at $z=0$, we then define the differences in the enclosed spherical masses of the progenitors 
within the neighboring redshifts of $z_1$ and $z_2>z_1$ as
\beqa
\Delta M_{\rm pseudo}(z_1) \equiv M(<R_{{\rm vir}, 1}, z_1) - M(<R_{{\rm vir}, 2}, z_1), \label{eq:delta_M_pseudo}
\eeqa
where $M(<r, z)$ represents the enclosed spherical mass of the progenitor at redshift $z$ within a sphere of $r$,
and $R_{{\rm vir}, j}$ is the virial radius with respect to $\Delta_{\rm vir}(z_j)$ times $\rho_{c}(z_j)$.
Eq.~(\ref{eq:delta_M_pseudo}) defines the redshift evolution of an SO mass between $z_1$ and $z_2$, 
when the density profile is assumed to be static during the finite redshift range.
It can be seen that Eq.~(\ref{eq:delta_M_pseudo}) corresponds to 
the first right-hand side term in Eq.~(\ref{eq:evol_SO_mass})
when performing a Taylor expansion of 
$M(<R_{{\rm vir}, 2}, z_1) \simeq M(<R_{{\rm vir}, 1}, z_1) + 4 \pi R_{{\rm vir}, 1}^2 \times \rho(R_{{\rm vir}, 1}, z_1) (R_{{\rm vir}, 2}-R_{{\rm vir}, 1})$. To find the virial radius $R_{\rm vir}$ from a binned enclosed mass profile, we use linear interpolation between two bins in logarithmic space.
Finally, we estimate the cumulative amount of pseudo evolution in the SO mass between 
$z=0$ and $z=1$ as\footnote{\ms{Note that $f_{\rm pseudo}=1$ does not correspond to perfectly stable halo density profiles between a finite redshift interval in our definition. This is simply because of the existence of mass accretion onto halos at large radii beyond the SO radii of interest.}}
\beqa
f_{\rm pseudo}(0<z<1) \equiv \frac{\sum_{0<z_{1}<1} \Delta M_{\rm pseudo}(z_1)}{M_{\rm vir}(z=0)-M_{\rm vir}(z=1)}, \label{eq:f_pseudo}
\eeqa
where $M_{\rm vir}(z)$ contains unbounded particles in the above equation.
It is worth noting that 29 snapshots are available between $z=0$ and $z=1$ in our simulations,
leading to a typical redshift width between two neighboring snapshots of $\sim0.03-0.04$.

\ms{Our estimation of Eq.~(\ref{eq:f_pseudo}) assumes no redshift evolution in the density profile between a small redshift interval. We evaluate possible systematic errors caused by the evolution between two snapshots in Appendix~\ref{app:sys_err_f_pseudo} and show that our estimates can have a $3-4$\% 
precision for the amount of pseudo evolution between $z=0$ and $z=1$.
}

\ms{Although we use the virial SO mass to study the amount of pseudo evolution throughout this study,
it would be interesting to consider the amount of pseudo evolution using different definitions for mass.
In Appendix~\ref{app:f_pseudo_m200b}, we examine the amount of pseudo evolution with an SO mass of $M_{\rm 200b}$. 
We find that the amount of pseudo evolution depends on the mass definition.
To be specific, the width in the distribution of $f_{\rm pseudo}(0<z<1)$ becomes narrower 
and the fraction of halos at $f_{\rm pseudo} > 0.75$ is reduced if the halo mass is defined by $M_{\rm 200b}$.
This is partly because the mass density profile within the radius of $R_{\rm 200b}$ is more universal across different redshifts than the counterparts within the virial radius on average \citep{2014ApJ...789....1D}.
These results are in good agreement with the findings in \citet{2013ApJ...766...25D}.}
 
\subsection{Age and density environments of dark matter halos} \label{subsec:age_and_tilde}

%According to its definition, our estimation of the amount of pseudo evolution (Eq.~[\ref{eq:f_pseudo}])
%is expected to be correlated with the age of a dark matter halo. 
\ms{Age is an important quantity to characterize the mass accretion history of a dark matter halo.}
To assess the correlation between age and the amount of pseudo evolution, we introduce
two common indicators, halo concentration $c_{\rm vir}$ and the half-mass scale factor $a_{1/2}$.
The former is defined by $R_{\rm vir}/r_s$ where $r_{s}$ is the scale radius of a universal density profile as proposed in \citet{Navarroetal:97}, while the latter is given by the scale factor $a=1/(1+z)$ 
at which the mass of the progenitor is equal to half of its descendant halo mass at $z=0$.
Previous numerical studies have suggested that the halo concentration at a fixed redshift 
depends not only on the halo mass, but also on the mass assembly history \citep[e.g.][]{2001MNRAS.321..559B, 2002ApJ...568...52W}.
In this study, we adopt the estimated value of $r_s$ for individual halos by the {\tt ROCKSTAR}
and compute $a_{1/2}$ from the merger tree.

\begin{figure*}[t!]
%\plottwo{rapid_acc_halo_projaxis0_Lproj20_mesh.pdf}{slow_acc_halo_projaxis1_Lproj20_mesh.pdf}
\gridline{\fig{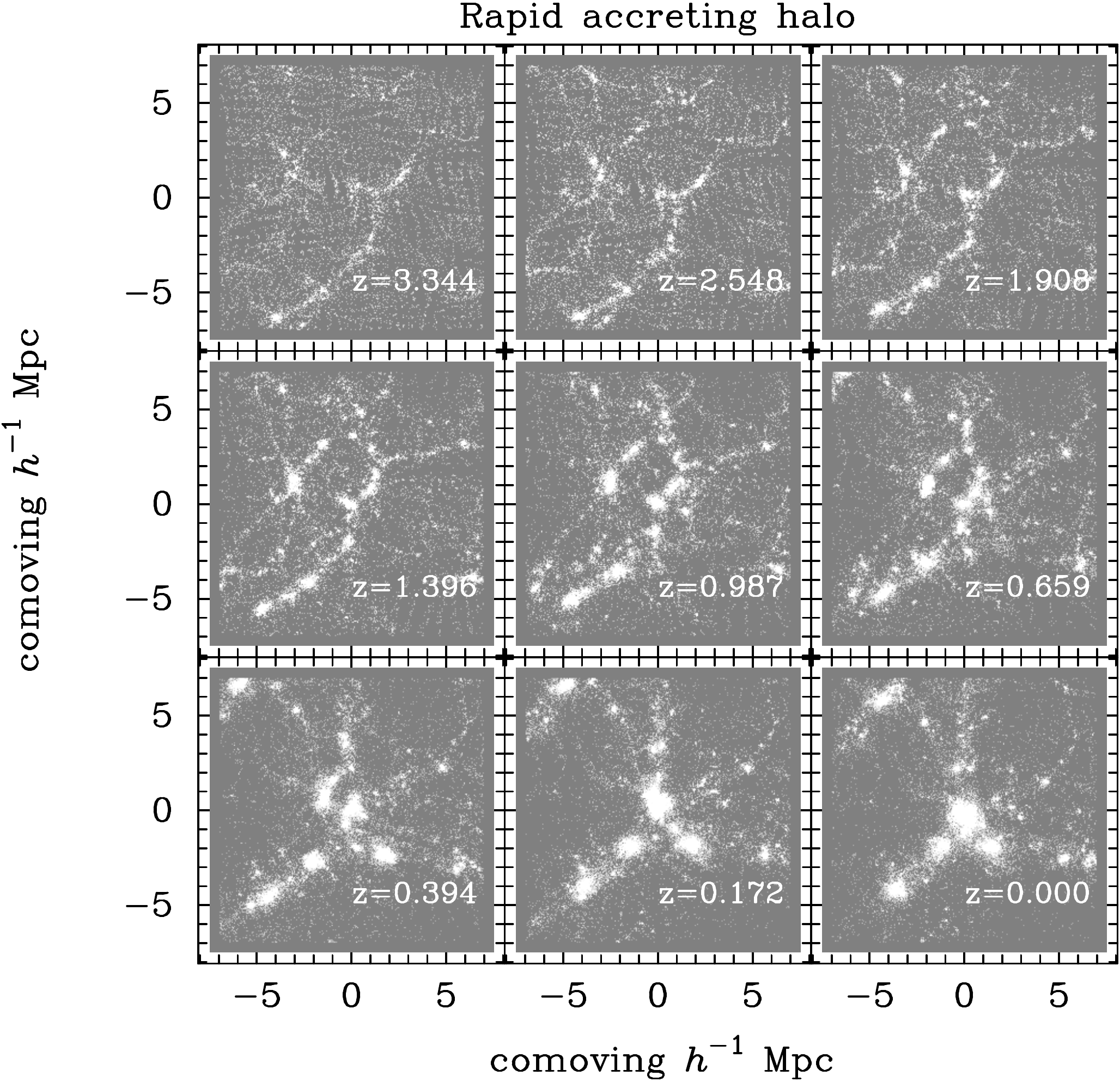}{0.47\textwidth}{}
          \fig{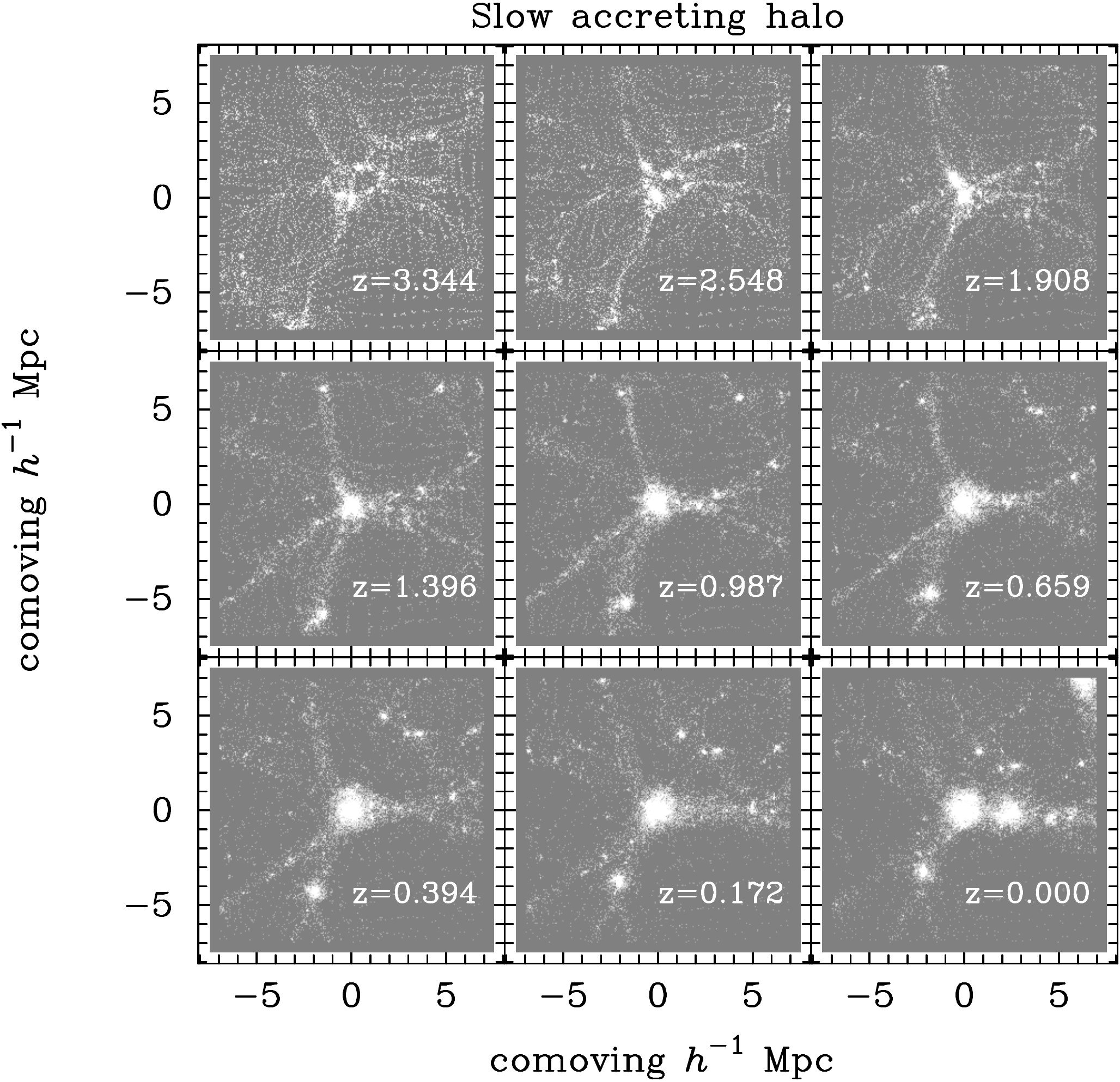}{0.47\textwidth}{}
          }
\caption{An example of mass accretion around galaxy clusters.
In this figure, we consider two simulated clusters containing a mass of $10^{14}\, h^{-1}\, M_{\odot}$
within the virial radius. Note that the corresponding virial radius is approximately $1 \, h^{-1}\, {\rm Mpc}$ at a redshift of $z=0$. The left panels show the distribution of projected mass density around a cluster with
a pseudo mass evolution fraction, from $z=1$ to 0, of $f_{\rm pseudo}(0<z<1) = 0.20$ 
(i.e. $80\%$ of the difference in virial mass between $z=1$ and 0 
can be accounted for by the physical mass accretion). 
The right panels represent a slowly accreting cluster with $f_{\rm pseudo}(0<z<1) = 0.77$.
The small panels in the left and right show the projected density at different redshifts from $z=3.344$ to $0$.
The cluster and its progenitor are located at the center of each image and the projection depth is set to be 
$2 \, h^{-1}\, {\rm Mpc}$. See the definition of $f_{\rm pseudo}(0<z<1)$ in Section~\ref{subsec:stats}.
\label{fig:example_acc_halos}}
\end{figure*}

In addition to the age, the amount of pseudo evolution can depend on the environment surrounding a dark matter halo,
because its exact value should closely relate to the mass accretion from outside the virial region
(e.g. a slow accreting halo would have a more stable density profile, leading to a greater amount of pseudo evolution of the SO mass in a finite redshift range). The dynamics of mass accretion onto a dark matter halo
will be regulated by the tidal tensor \citep[e.g.][]{2009MNRAS.398.1742H},
\beqa
T_{ij} \equiv \left[ \frac{\partial^{2}}{\partial_{i}\partial_{j}}-\frac{1}{3}\delta^{\rm K}_{ij} \nabla^2 \right]\phi, 
\eeqa
where $\phi$ is the gravitational potential and $\delta^{\rm K}_{ij}$ represents the Kronecker symbol.
The tidal shear $q^2$ is a quantity that is promising for the characterization of 
the local environment of halos \citep{1988MNRAS.232..339H, 1996MNRAS.282..436C}, 
which is commonly defined by 
\beqa
q^2 \equiv \frac{1}{2}\left[
\left(\lambda_{3}-\lambda_{1}\right)^2
+\left(\lambda_{3}-\lambda_{2}\right)^2
+\left(\lambda_{2}-\lambda_{1}\right)^2
\right],
\eeqa
where $\lambda_{1} \le \lambda_{2} \le \lambda_{3}$ are the eigenvalues of $T_{ij}$.
The previous numerical studies have clearly demonstrated that tidal shear plays an important role in determining
not only the dynamical properties, but also the large-scale clustering of halos \citep[e.g.][]{2011MNRAS.413.1973W, 2012PhRvD..86h3540B, 2015ApJ...807...37S, 2018MNRAS.476.3631P}. Motivated by these results, we consider 
the tidal tensor induced by the mass density distribution surrounding a dark matter halo.
To be specific, we compute an external tidal shear from the distribution of $N$-body particles as
\beqa
T^{\rm ext}_{ij} \equiv \sum_{q} 
\frac{G m_{p}}{r^{5}_{q}}\left(3 r_{q,i}r_{q,j} - \delta^{\rm K}_{ij} r^{2}_{q}\right), \label{eq:ext_tilde}
\eeqa
where $m_{p}$ is the mass of a particle,
$\bd{r}_{q}$ is a vector locating the $q$-th particle from the center of a halo,
and the summation runs over if $r_q$ ranges between $r_{\rm min}$ and $r_{\rm max}$.
Throughout this study, we set $r_{\rm min} = R_{\rm 200b}$ and $r_{\rm max}=10\, R_{\rm 200b}$
when computing Eq.~(\ref{eq:ext_tilde}) for individual halos at $z=0$.
We then compute the tidal shear $(q^{\rm ext})^2$ from the eigenvalues of $T^{\rm ext}_{ij}$
and cancel a trivial correlation between the tidal shear and the amplitude of over-density as follows:
\beqa
\alpha = \frac{q^{\rm ext}}{1+\delta^{\rm ext}}, \, 
1+\delta^{\rm ext} \equiv
\frac{\int_{r_{\rm min}}^{r_{\rm max}}\, 4\pi r^2{\rm d}r \rho(r, z=0)}{4\pi/3\, \bar{\rho}_{\rm m0}\left( r^{3}_{\rm max}-r^3_{\rm min}\right)}, \label{eq:reduced_shear}
\eeqa
where $\bar{\rho}_{\rm m0} = \bar{\rho}_{\rm m}(z=0)$. 
We denote the reduced tidal shear as $\alpha$ in this study.
Note that \citet{2018MNRAS.476.3631P} introduced a similar quantity for tidal anisotropy defined
from the smoothed density field around individual halos with a Gaussian smoothing scale of $4 R_{\rm 200b}$
and showed that tidal anisotropy strongly correlates with halo bias. As shown in later parts of the paper, 
we also find a strong correlation between the reduced tidal shear $\alpha$ and the large-scale halo bias.

\section{Results} \label{sec:res}

\subsection{Visual impression} \label{subsec:visual}

We first make a visual comparison between two dark matter halos with a virial mass 
of $M_{\rm vir}=10^{14}\, h^{-1}M_{\odot}$ at $z=0$, but with a different amount of pseudo evolution in 
$M_{\rm vir}(z)$ between $z=0$ and $1$.
Figure~\ref{fig:example_acc_halos} shows the mass accretion history of the two halos:
the left represents the history of the halo with $f_{\rm pseudo}(0<z<1) = 0.20$,
while the right is for the halo with $f_{\rm pseudo}(0<z<1) = 0.77$.
For each, the small panel reveals the projected mass density with a projection depth of 
$2\, h^{-1}{\rm Mpc}$ at different redshifts. The halo and its progenitor are located at 
the center of each small panel. Note that the virial radius is found to be $1\, h^{-1}{\rm Mpc}$ at $z=0$ 
in both panels.

\begin{figure*}[t!]
%\plotone{hist_frac_pseudo_z0.pdf}
\gridline{\fig{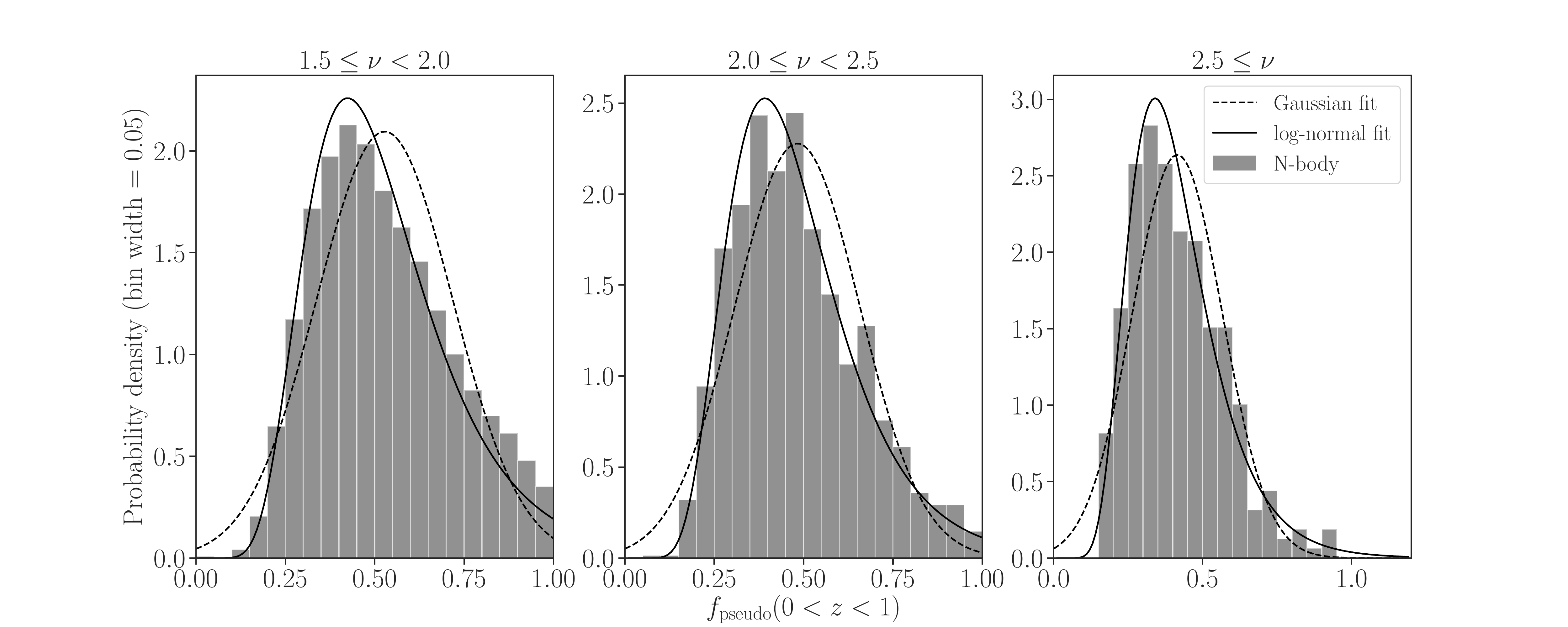}{1.\textwidth}{}}
\caption{The probability distribution function (PDF) of the fraction of pseudo mass evolution from $z=1$ to 0, denoted as $f_{\rm pseudo}(0<z<1)$. The left, middle, and right panels show 
the PDFs for clusters with different ranges in peak height $\nu=\delta_{\rm c}/\sigma$, 
%\nu=\delta_{\rm c}/\sigma=1.5-2$, $2.0-2.5$ and larger than $2.5$, 
$1.5 \le \nu < 2$, $2 \le \nu < 2.5$, and $\nu \ge 2.5$, respectively. 
The gray histogram represents the result from the $N$-body simulations,
while the dashed and solid lines show Gaussian and log-normal fits.
\label{fig:hist_frac_pseudo_z0}}
\end{figure*}

\begin{table*}[ht!]
\renewcommand{\thetable}{\arabic{table}}
\centering
\caption{Mean and root-mean square (RMS) of the amount of pseudo evolution of virial SO mass.} \label{tab:table_f_mean_std}
\begin{tabular}{c|cc|cc}
\tablewidth{0pt}
\hline
\hline
Range of $\nu$ & Mean (linear) & RMS (linear) & Mean ($10$-base $\log$) & RMS ($10$-base $\log$) \\
\hline
\decimals
$1.5 \le \nu < 2.0$ & \ms{$0.527$} & \ms{$0.190$} & \ms{$-0.307$} & \ms{$0.167$} \\
$2.0 \le \nu < 2.5$ & \ms{$0.482$} & \ms{$0.175$} & \ms{$-0.346$} & \ms{$0.163$} \\
$\nu \ge 2.5$       & \ms{$0.415$} & \ms{$0.151$} & \ms{$-0.410$} & \ms{$0.158$} \\
\hline
\end{tabular}
\end{table*}

The figure illustrates the rich large-scale structures that occur around a cluster-sized dark matter halo.
Such structures may be associated with the mass accretion history.
In the left, the halo has a small amount of pseudo evolution and this indicates 
that rapid growth has occurred in the mass density profile between $z=0$ and $1$.
As shown in the figure, the clumpier and richer structures found beyond the virial regime are more prevalent
in the left as compared to the right.
Those structures introduce an efficient mass accretion 
onto the cluster-sized dark matter halo at the center of each small panel.
Therefore, we expect the amount of pseudo evolution of an SO mass to be correlated with
the environment of an individual halo \ms{\citep[also, see][for the relation between the outer mass profile and the physical mass accretion]{2016ApJ...818..188D}}. 

% nu = 1.5-2 (mean mass M200b = 8.0e13 Msun/h)
% f<0 f>1 total
% 5 735 10525
% Gaussian fit: mean = 0.5278048567233521, rms = 0.19041769581435145
% log10 normal: mean = -0.3077058327584342, rms = 0.16779688952048324

% nu = 2-2.5 (mean mass M200b = 2.6e14 Msun/h)
% f<0 f>1 total
% 0 45 1549
% Gaussian fit: mean = 0.48250295139627664, rms = 0.17514431950720888
% log10 normal: mean = -0.34593481488975214, rms = 0.1632080028598134

% nu > 2.5 (mean mass M200b = 7.1e14 Msun/h)
% f<0 f>1 total
% 1 1 320
% Gaussian fit: mean = 0.4148072339622641, rms = 0.15130042283287337
% log10 normal: mean = -0.4106209862278978, rms = 0.15852045605104703

%% The "ht!" tells LaTeX to put the figure "here" first, at the "top" next
%% and to override the normal way of calculating a float position

\subsection{Amount of pseudo mass evolution} \label{subsec:f_pseudo}

We explore the statistical property of the amount of pseudo evolution $f_{\rm pseudo}(0<z<1)$ as defined by Eq.~(\ref{eq:f_pseudo}).
To study the halo mass dependence, 
we divide the halo sample into three sub-samples based on peak height $\nu=\delta_{\rm c}/\sigma$.
The bin of the peak height is set to $1.5 \le \nu<2$, $2 \le \nu<2.5$, and $\nu \ge 2.5$ at $z=0$.
After selection, we find the available number of halos are 10525, 1549, and 320 
for the three $\nu$ bins, respectively.
The mean halo masses at $z=0$ in the three sub-samples are 
found to be $M_{\rm 200b}=8.0\times10^{13}\, h^{-1}M_{\odot}$,
$2.6\times10^{14}\, h^{-1}M_{\odot}$, and $7.1\times10^{14}\, h^{-1}M_{\odot}$.
In the following, we limit the range of $f_{\rm pseudo}(0<z<1)$ 
for individual halos to be between 0 and 1. 
\ms{The number of halos within our halo sample with $f_{\rm pseudo}>1$ is 
found to be 735, 45, and 1 for the samples with $1.5 \le \nu<2$, $2 \le \nu<2.5$, and $\nu\ge 2.5$, respectively.}
These minor populations are expected to have undergone tidal stripping, because $f_{\rm pseudo}>1$
indicates that the density profile around the virial radius decreases as $z=1$.
\ms{We also found a halo with $f_{\rm pseudo}<0$ in our sample and this ill-defined value is
caused by the major merger at $z=0.1$ and the significant change in its density profile after $z=0.1$.}
Our estimation of the amount of pseudo evolution implicitly assumes that the density profile would evolve smoothly as a function of redshift. Therefore, we simply remove the halo with $f_{\rm pseudo}<0$ in this study.
\ms{The measurements of $f_{\rm pseudo}$ for five halos were also not included because of a lack of information concerning the merger tree.}

\begin{figure*}[ht!]
%\plotone{scatter_plot_halo_prop_z0.pdf}
\gridline{\fig{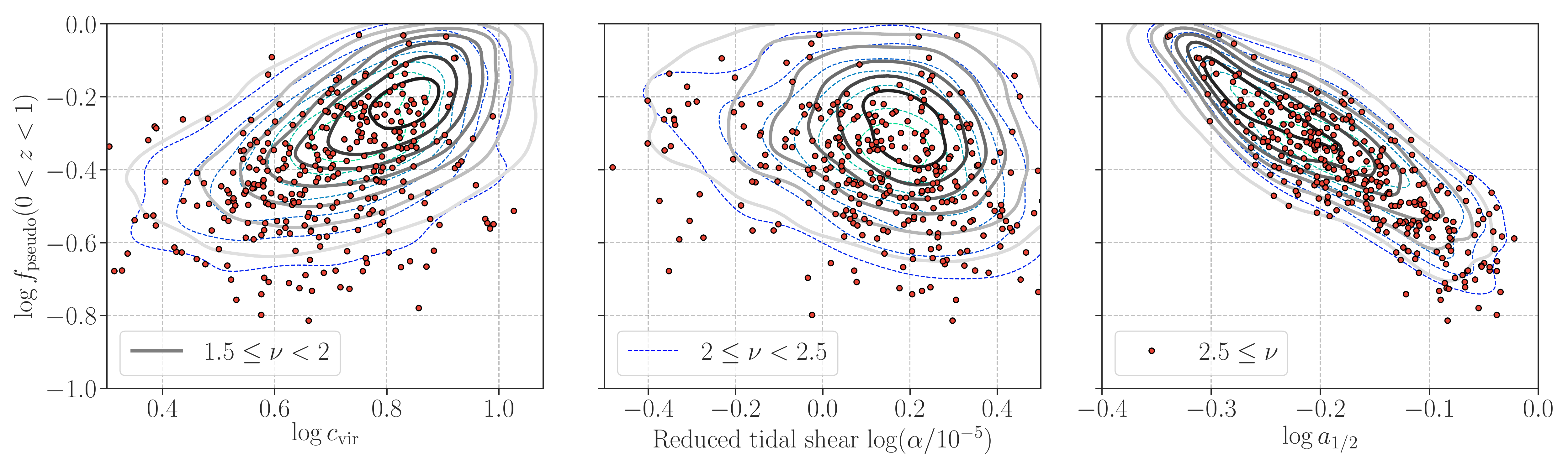}{1.\textwidth}{}}
\caption{Scatter plot between the fraction of pseudo mass evolution from $z=1$ to 0 ($f_{\rm pseudo}(0<z<1)$)
and other halo properties. The left, middle, and right panels show the correlation between $f_{\rm pseudo}(0<z<1)$ and concentration, reduced tidal shear $\alpha$ and half-mass scale factor $a_{1/2}$, respectively. In each panel, the solid contours represent the results for the clusters in the peak height range $1.5<\nu<2$, the dashed contours represent clusters with $2<\nu<2.5$, and the red points are 
clusters with $\nu>2.5$. \label{fig:scatter_plot_prop_z0}}
\end{figure*}

\begin{table*}[ht!]
\renewcommand{\thetable}{\arabic{table}}
\centering
\caption{Correlation of the amount of pseudo evolution of virial SO mass with other halo properties.
We denote the halo concentration, the half-mass scale factor, and the reduced tidal shear
as $c_{\rm vir}$, $a_{1/2}$ and $\alpha$, respectively. See Eq.~(\ref{eq:reduced_shear}) for 
our definition of $\alpha$. Note that $\log$ represents a 10-base logarithm. } \label{tab:table_f_correlation}
\begin{tabular}{c|ccc}
\tablewidth{0pt}
\hline
\hline
Range of $\nu$ 
&
$\log c_{\rm vir}$ -- $\log f_{\rm pseudo}$ 
&
$\log (\alpha/10^{-5})$ -- $\log f_{\rm pseudo}$
& 
$\log a_{1/2}$ -- $\log f_{\rm pseudo}$
\\
\hline
\decimals
$1.5 \le \nu < 2.0$ & \ms{$0.319$} & \ms{$-0.336$} & \ms{$-0.828$} \\
$2.0 \le \nu < 2.5$ & \ms{$0.429$} & \ms{$-0.253$} & \ms{$-0.810$} \\
$\nu \ge 2.5$       & \ms{$0.400$} & \ms{$-0.142$} & \ms{$-0.760$} \\
\hline
\end{tabular}
\end{table*}

Figure~\ref{fig:hist_frac_pseudo_z0} shows the histogram of $f_{\rm pseudo}(0<z<1)$ for our three subsamples 
at $z=0$. In each panel, the gray boxes represent the results of the simulation, while the solid and dashed lines
are the Gaussian and log-normal fits, respectively. 
The fitting results are provided in Table~\ref{tab:table_f_mean_std}.
\ms{The relationship between averaged $f_{\rm pseudo}(0<z<1)$ and halo mass is given by
\beqa
\langle f_{\rm pseudo}(0<z<1) \rangle &=& 
(-0.0193\pm0.0013) M_{\rm 200b, 14}  \nonumber \\
&& + (0.5421\pm0.0024), \label{eq:f_mean}
\eeqa
where 
$M_{\rm 200b, 14} =\left( M_{\rm 200b}/10^{14}\, h^{-1}M_{\odot} \right)$,
and 
we find a clear decreasing trend in $f_{\rm pseudo}$ as a function of halo mass.
The variance around Eq.~(\ref{eq:f_mean}) is approximated as 
$(-0.0022\pm0.0003) M_{\rm 200b, 14} + (0.0379\pm0.0006)$, 
suggesting that more massive galaxy clusters have smaller scatters in $f_{\rm pseudo}$.}

% correlation coefficient in log c vs log f
% nu = 1.5-2 : 0.31977165
% nu = 2-2.5 : 0.42947626
% nu = 2.5- : 0.40029082

% correlation coefficient in log alpha/1e-5 vs log f
% nu = 1.5-2 : -0.33603798
% nu = 2-2.5 : -0.25316372
% nu = 2.5- : -0.14240916

% correlation coefficient in log a_{1/2} vs log f
% nu = 1.5-2 : -0.82852304
% nu = 2-2.5 : -0.81072608
% nu = 2.5- : -0.76029977

Next, we study the correlation between the amount of pseudo evolution and other halo properties. Figure~\ref{fig:scatter_plot_prop_z0} summarizes
the correlations measured in our simulation. 
In the figure, we consider three halo properties apart from the amount of pseudo evolution;
the halo concentration $c_{\rm vir}$, the half-mass time $a_{1/2}$,
and the reduced tidal shear defined in Eq.~(\ref{eq:reduced_shear}).
In each panel of Figure~\ref{fig:scatter_plot_prop_z0}, the thick and dashed contour lines show the correlations for halos with $1.5 \le \nu<2$ and $2 \le \nu<2.5$, while the red points illustrate halos with $\nu \ge 2.5$.
We find that the amount of pseudo evolution shows a positive correlation 
with $c_{\rm vir}$, while it is anti-correlated with the density of the environment beyond the virial scales $\alpha$, and the halo age $a_{1/2}$.
The positive correlation between $f_{\rm pseudo}$ and $c_{\rm vir}$
is expected \ms{because older halos are expected to include more stable density profiles , 
i.e. larger $f_{\rm pseudo}$, with greater concentration \citep[e.g.][]{2001MNRAS.321..559B, 2002ApJ...568...52W}.}
Conversely, a weak anti-correlation between $f_{\rm pseudo}$
and a reduced tidal shear $\alpha$ may be understood via the following.
A density environment with larger $\alpha$ 
indicates the presence of richer filamentary structures around a halo.
Such filaments can induce an efficient mass accretion onto the halo
from the scales beyond the virial regime, implying that physical mass accretion can be a main driver in the time evolution of virial SO mass,
i.e. smaller $f_{\rm pseudo}$.
This argument is supported by the visual comparison shown in Figure~\ref{fig:example_acc_halos}.
It is also worth noting that our estimation of tidal shear is mainly determined from the density structures around the virial radius (see Eq.~[\ref{eq:ext_tilde}]).
In fact, we found no correlations of $f_{\rm pseudo}$ between
the tidal tensors of a smoothed density field with a Gaussian smoothing scale 
of $10 \, h^{-1}{\rm Mpc}$\footnote{To be specific, we use the quantity $\alpha_{R}$ defined in Eq.~(10) in \citet{2018MNRAS.476.3631P} for an indicator of tildes. The correlation of $f_{\rm pseudo}$ with $\alpha_{R}$ is found to be less than 0.10 for $\nu\ge1.5$ when adopting a smoothing scale of $10 \, h^{-1}{\rm Mpc}$.}.
Hence, we expect the amount of pseudo evolution to depend on the density environment close to the virial radius, but not on the large-scale environment.
The cross correlation coefficients in logarithmic space are summarized in Table~\ref{tab:table_f_correlation}.

\begin{figure*}[t!]
\gridline{\fig{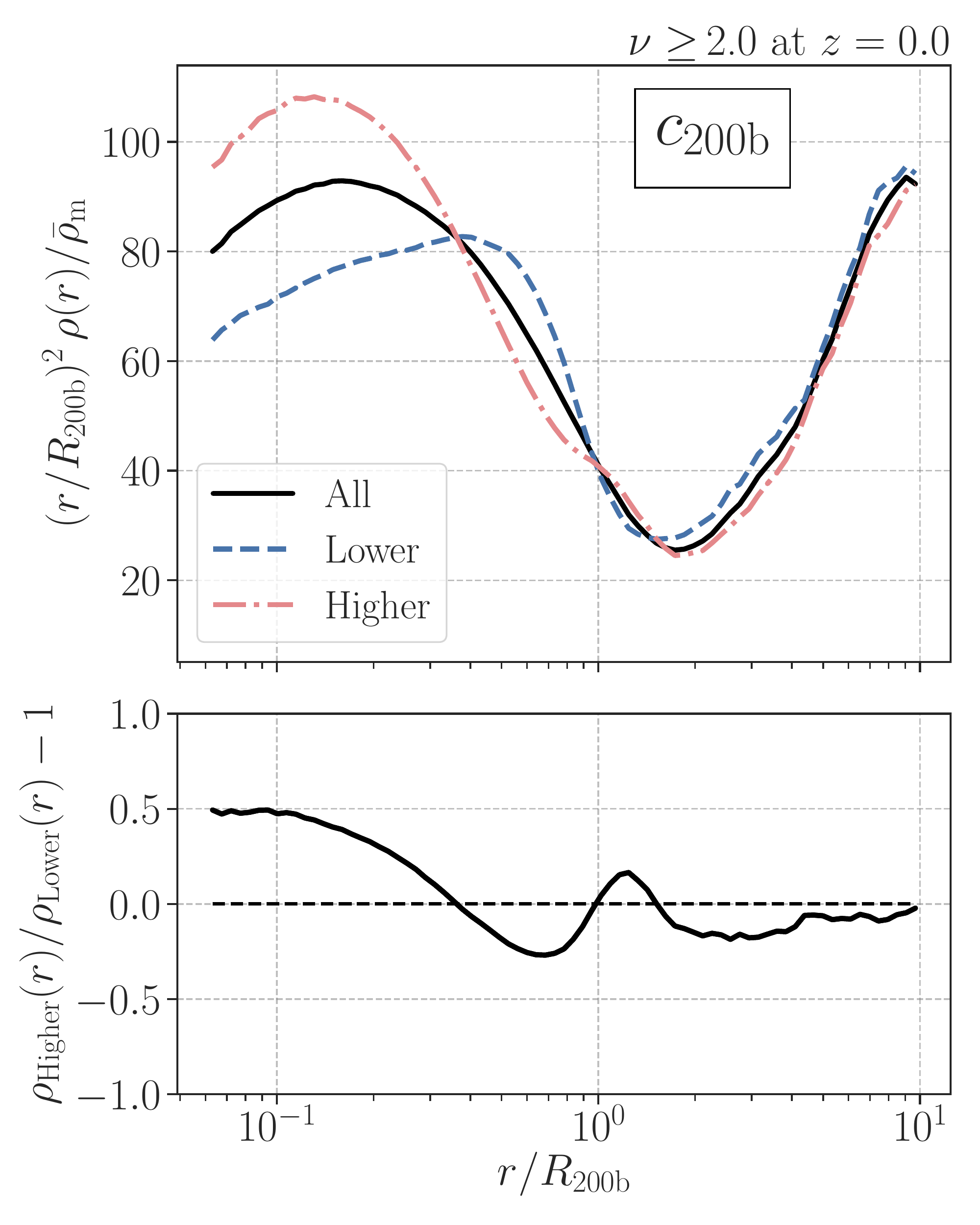}{0.34\textwidth}{(a)}
          \fig{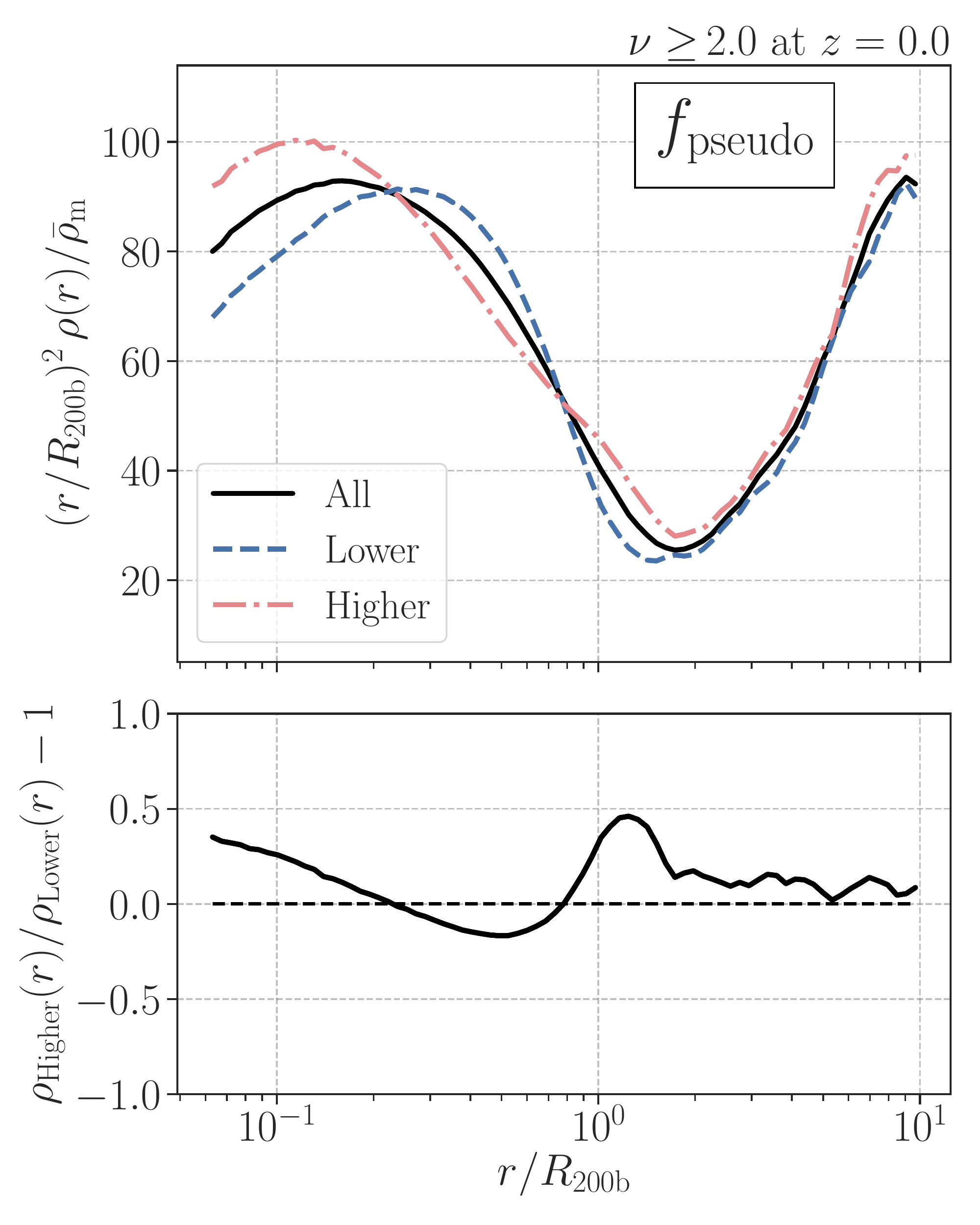}{0.34\textwidth}{(b)}
          \fig{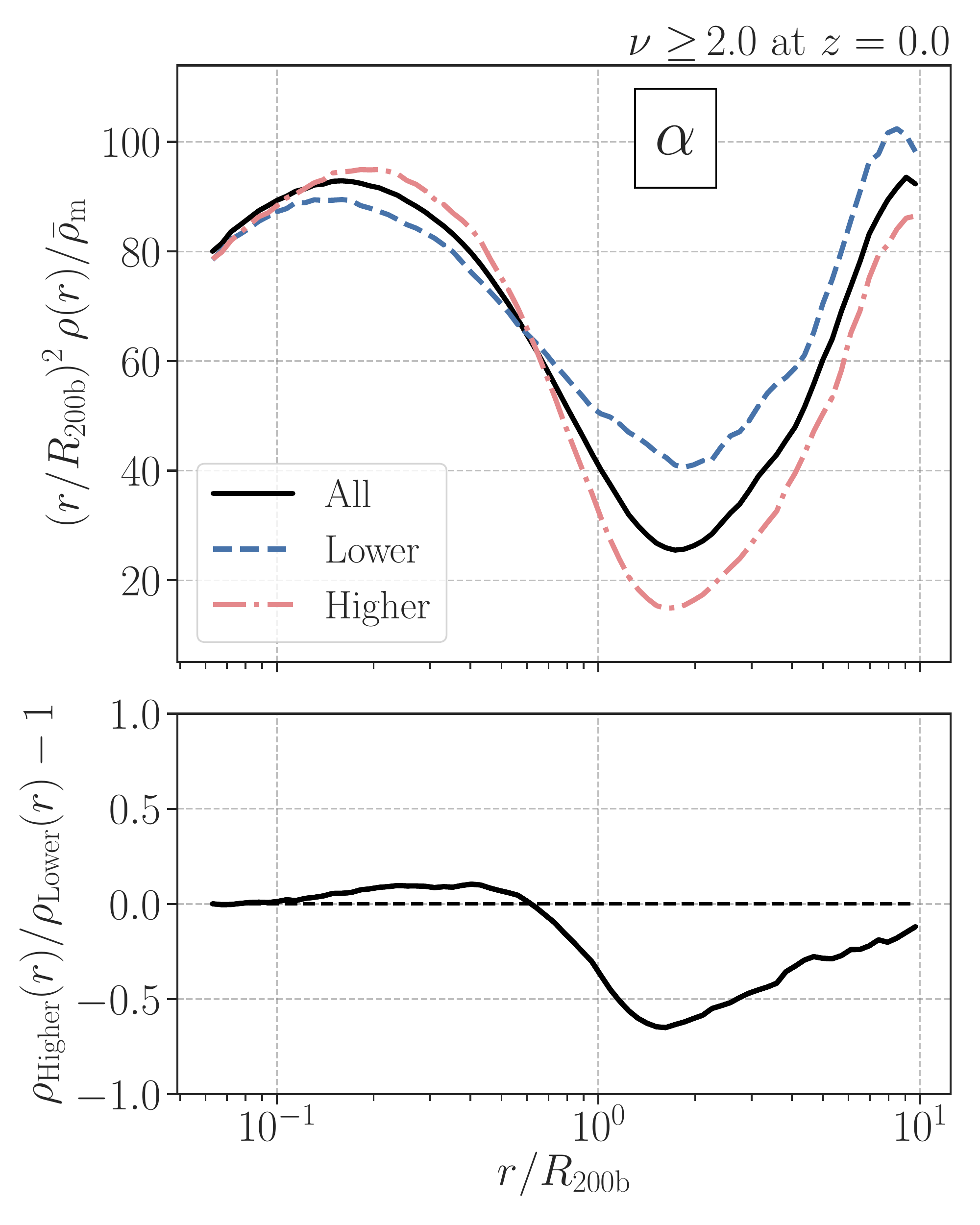}{0.34\textwidth}{(c)}
          }
\caption{
Stacked mass density profile of cluster-sized dark matter halos with a peak height parameter larger than $2$.
In each top panel, the solid line shows the result for a full sample, while the dashed and dotted-dashed lines
show the stacked profiles for subsamples with $33\%$ lower and higher values of a secondary halo parameter, respectively.
In the bottom panels, we show the fractional difference in the stacked density profiles between two subsamples.
In this figure, we consider three secondary parameters:
(a) halo concentration \ms{$c_{\rm 200b}$}, (b) the fraction of pseudo mass evolution from $z=1$ to 0 and (c) reduced tidal shear $\alpha$. Note that the tidal shear is computed from $N$-body particles in a spherical shell with an inner radius of $R_{\rm 200b}$ and an outer radius of $10\, R_{\rm 200b}$, where $R_{\rm 200b}$ is the spherical over-density radius of each halo of 200 times the mean matter density.
\label{fig:stack_prof}}
\end{figure*}

\subsection{Stacked density profile} \label{subsec:stack_prof}

We then study the \ms{connection} between the pseudo evolution in cluster SO masses
\ms{to} and spherical density profiles and the amplitude of large-scale clustering.
For this purpose, we perform a stacking analysis of spherical density profiles for a set of halos. 
We here focus on halos with $\nu \ge 2$ 
at $z=0$. Note that the average halo mass is found to be 
$3.3\times10^{14}\, h^{-1}M_{\odot}$,
which is a typical mass of a galaxy cluster at low redshift.
Each halo has the scale radius 
\ms{$r_{s}$}, the reduced tidal shear $\alpha$, and 
the amount of pseudo evolution in its virial SO mass of between $z=0$ and $1$, 
$f_{\rm pseudo}(0<z<1)$. We divide the parent halo sample into three by
using a secondary parameter (\ms{$c_{\rm 200b}=R_{\rm 200b}/r_s$}, $\alpha$, and $f_{\rm pseudo}$)
so that the average halo mass and 
the number of each subsamples will be equal.
After dividing the halo sample, we find that the average masses of the halos within the three subsamples match within the level of 
a few percent.
We then stack the density profile for the parent samples 
and the subsamples with 33\% lower and higher values of a secondary parameter. 

Figure~\ref{fig:stack_prof} shows the results of the stacking analysis of the spherical density profiles. 
From left to right, we consider 
the subsamples defined by \ms{$c_{\rm 200b}$}, $f_{\rm pseudo}$, and $\alpha$.
In each top panel, the black solid line shows the stacked profile of the parent halo sample, while the blue dashed and red dotted-dashed lines
are for the lower and higher subsamples.
In each bottom panel, we show the fractional difference 
in the stacked profiles between the higher and lower subsamples.
Note that the top panels of Figure~\ref{fig:stack_prof} are given in the form $\sim r^2 \rho(r)$ to allow the easy location of the scale radius of the stacked density profiles where ${\rm d}\ln\rho(r)/{\rm d}\ln r = -2$. 

First we confirm that the stacked density profile of those halos with lower concentrations
shows a larger amplitude outside of the virial radius compared to the full sample, as in previous studies \citep[e.g.][]{2006ApJ...652...71W}.
Since the amplitude of a stacked density profile outside a 
virial regime is closely associated with the linear halo bias \citep[e.g.][]{2008MNRAS.388....2H}, 
the top left panel in Figure~\ref{fig:stack_prof} 
represents 
the subsample with a lower \ms{$c_{\rm 200b}$}, which has the larger linear halo bias
and is consistent with the known secondary halo bias on galaxy-cluster scales \citep[e.g.][]{2008ApJ...687...12D}.
For subsamples with $f_{\rm pseudo}$, the relationship between halo concentration and linear bias is found to be inverted.
The top middle panel in Figure~\ref{fig:stack_prof} 
shows that the lower-$f_{\rm pseudo}$ subsample can have 
a lower concentration ${\it and}$ and a lower linear bias.
As shown in Figure~\ref{fig:scatter_plot_prop_z0}, the amount of pseudo evolution $f_{\rm pseudo}$ is correlated with the halo concentration.
Therefore, it is a natural consequence that a stacked density profile 
of halos with lower $f_{\rm pseudo}$ will have a lower \ms{$c_{\rm 200b}$}.
\ms{Note that our stacked analysis of density profiles split by halo property is found to be consistent with the results in \citet{2014ApJ...789....1D}. We confirmed that halos with high mass accretion rates exhibit very different median profiles compared to their slowly accreting counterparts, while lower-concentration samples have steeper outer density profiles. }

To understand why the lower-$f_{\rm pseudo}$ subsample can have the smaller linear bias,
we look into the stacked density profile of the subsamples with the reduced tidal shear $\alpha$.
The top right panel in Figure~\ref{fig:stack_prof} presents the stacked density profiles for the subsamples divided by $\alpha$.
It clearly shows that the reduced tidal shear $\alpha$ has only a limited effect on the inner density profile, i.e. the halo concentration.
On the other hand, the value of $\alpha$ 
is tightly anti-correlated with the linear halo bias.
Hence, Figure~\ref{fig:stack_prof} highlights that
the inner stacked density profile for halos with different $f_{\rm pseudo}$
provides the correlation between $f_{\rm pseudo}$ and \ms{$c_{\rm 200b}$} (or halo age), whereas the outer profile is probably governed by the anisotropy in the density of the environment.

\subsection{Linear bias and concentration}\label{subsec:bias_and_c200}

To quantify the results in Section~\ref{subsec:stack_prof},
we infer the linear bias and the halo concentration 
from statistical analyses of the halo samples.

\begin{figure}[t!]
\gridline{\fig{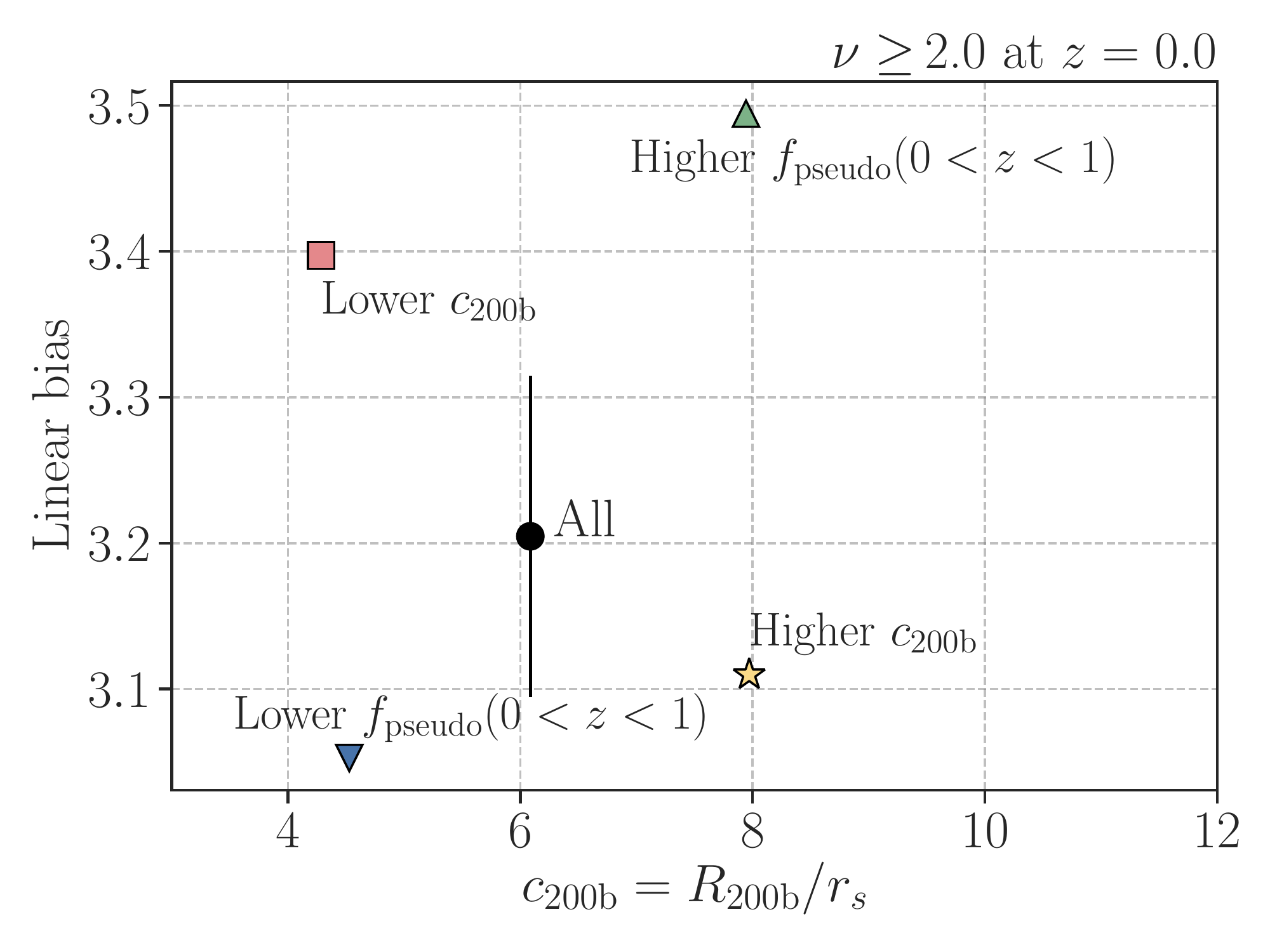}{0.4\textwidth}{}}
\caption{The relationship between linear bias and concentration for present-day clusters 
with a peak height larger than $2$. The black point shows the results for the full sample and the error bar is the Gaussian error of the estimation of the linear bias in the simulation. The different symbols represent the bias-concentration relationships of the subsample divided by the secondary parameters of concentration (\ms{$c_{\rm 200b}$}) 
or the amount of pseudo mass evolution, $f_{\rm pseudo}(0<z<1)$. \label{fig:c200b_bias}}
\end{figure}

We evaluate the linear halo bias by using 
an auto power spectrum of the matter density field $P_{\rm mm}(k)$,
and a cross power spectrum between matter and 
the halo overdensity fields $P_{\rm hm}(k)$.
To measure these power spectra, we construct matter and halo overdensity fields from $N$-body particles and halos using Clouds-in-Cell (CIC) assignment, respectively. 
For the CIC assignment, a $1024^3$ cubic lattice is adopted.
The power spectrum is computed in Fourier space with linear spaced bins
and a bin size of $\Delta k=0.012\, h \, {\rm Mpc}^{-1}$.
We then estimate the linear halo bias $b_{\rm L}$
so as to minimize the following $\chi^2$ statistic:
\beqa
\chi^{2}(b_{\rm L}) = \sum_{\, \, \, \, \, \, k_{i}\le k_{\rm max}}\frac{\left[P_{\rm hm}(k_{i})-b_{\rm L}P_{\rm mm}(k_{i})\right]^2}{\sigma^2_{P_{\rm hm}}(k_{i})}, \label{eq:chi2}
\eeqa
where we set $k_{\rm max}=0.1\, h \, {\rm Mpc}^{-1}$
and $\sigma_{P_{\rm hm}}$ is 
the Gaussian error of the cross power spectrum $P_{\rm hm}$.
The Gaussian error of $P_{\rm hm}$ is evaluated as
\beqa
\sigma^2_{P_{\rm hm}}(k_{i}) = 
\frac{P_{\rm hh}(k_{i})P_{\rm mm}(k_{i})+P^2_{\rm hm}(k_{i})}{N_{\rm mode}(k_{i})},
\eeqa
where $N_{\rm mode}(k_{i})$ is the number of Fourier modes 
in $i$-th bin, and $P_{\rm hh}$ is 
the auto power spectrum of the halo overdensity field including the shot noise.
We also estimate the uncertainty in $b_{\rm L}$ by 
imposing $\chi^2-\chi^2_{\rm min} \le 1$, where $\chi^2_{\rm min}$
is the minimum $\chi^2$ statistic.

We also re-measure the halo concentration from the stacked density profile in a radius range of $0.05\le r/R_{\rm 200b}\le 0.5$
by using the profile proposed in \citet{Navarroetal:97}\footnote{\ms{This re-fitting is needed because the concentration of a stacked profile is not generally the mean or median of the concentrations of the individual profiles.}},
\beqa
%\rho_{\rm Einasto}(x) &=& \rho_{s} \exp\left\{-\frac{2}{\alpha_{\rm E}}
%\left[\left(c_{\rm 200b}x \right)^{\alpha_{\rm E}}-1\right]\right\},
%\label{eq:Einasto}
\rho_{\rm NFW}(x) &=& \frac{\rho_{s}}{(c_{\rm 200b} x)\, (1+c_{\rm 200b}x)^2}, \label{eq:NFW}
\eeqa
where 
$x = r/R_{\rm 200b}$ and $c_{\rm 200b} = R_{\rm 200b}/r_s$.
%$\alpha_{\rm E}$ is the shape parameter and 
%we adopt the relation between $\alpha_{\rm E}$ and peak height $\nu$ as calibrated in %\citet{2008MNRAS.387..536G}.
the fit was carried out using 
the non-linear least-squares Levenberg-Marquardt algorithm \citep{1992nrfa.book.....P}
by minimizing the $\chi^2$-fitting metric defined as
\beqa
\chi^2_{\rho} = \sum_{i=1}^{N_{\rm bin}}\left[\ln \rho(x_{i}) - \ln \rho_{\rm NFW}(x_{i} | \rho_{s}, c_{\rm 200b}) \right]^2,
\eeqa
where $N_{\rm bin}$ is the number of radial bins in the stacked density profile. We find that $N_{\rm bin}=32$ when limiting the radial range to be $0.05\le r/R_{\rm 200b}\le 0.5$.

Figure~\ref{fig:c200b_bias} summarizes the fitting results 
of the linear halo bias $b_{\rm L}$ and 
the concentration $c_{\rm 200b}$ for different halo samples. 
The figure shows that 
\ms{the amount of pseudo evolution $f_{\rm pseudo}$ 
can change the linear halo bias by $6.25\%$.}
This difference in the linear bias with respect to $f_{\rm pseudo}$
is at a similar level to the known secondary bias due to the halo concentration.

\subsection{Varying halo masses and redshifts} \label{subsec:diff_m_z}

So far, we have focused on halos at $z=0$ and 
the amount of pseudo evolution that has taken place in the virial SO mass since $z=1$. 
In this section, we generalize the aforementioned results 
by considering the different redshift ranges.
We therefore consider three halo samples 
at different redshifts $z=0.3$, $0.6$, and $1$,
and estimate the amount of pseudo evolution since $z=1.6$, $2$, and $3$ 
as in Section~\ref{subsec:stats}, respectively.
We denote the amount of pseudo evolution from $z=z_{\rm i}$ to $z_{\rm f}$
as $f_{\rm pseudo}(z_{\rm f} < z < z_{\rm i})$.
\ms{For a given $z_{\rm f}$, we determine the initial redshift $z_{\rm i}$ by imposing
$t(z_{\rm i}) = t(z_{\rm f}) - 2 t_{\rm dyn}(z_{\rm f})$, 
where $t(z)$ represents the cosmic time at redshift $z$ and $t_{\rm dyn}(z)$ 
is the dynamical time for the SO mass with respect to 200 times $\bar{\rho}_{\rm m}$.
In our adopted cosmological model, this criteria sets $z_{\rm i}=1, 1.6, 2$ and $3$ 
for $z_{\rm f}=0, 0.3, 0.6$, and $2$, respectively.}
%As follows in Section~\ref{subsec:stack_prof},
%\ms{we set the minimum halo mass at each redshift by $\delta_{c}/\sigma(M_{\rm 200b}, z)\ge2$}
%, where $\sigma(M,z)$ is the mass variance derived by linear matter power spectrum at $z$ with a spherical top-hat filter.
%Note that the top-hat radius is set in the same way as in 
%Section~\ref{subsec:catalog}.
We then perform the stacking analysis of the spherical density profiles
for halo catalogs at $z=0.3$, $0.6$, and $1$ 
with the amount of pseudo evolution used for the selection.
We select the halo samples with 33\% higher and lower $f_{\rm pseudo}$
at different redshifts as in Section~\ref{subsec:stack_prof}.

\begin{figure}[t!]
\gridline{\fig{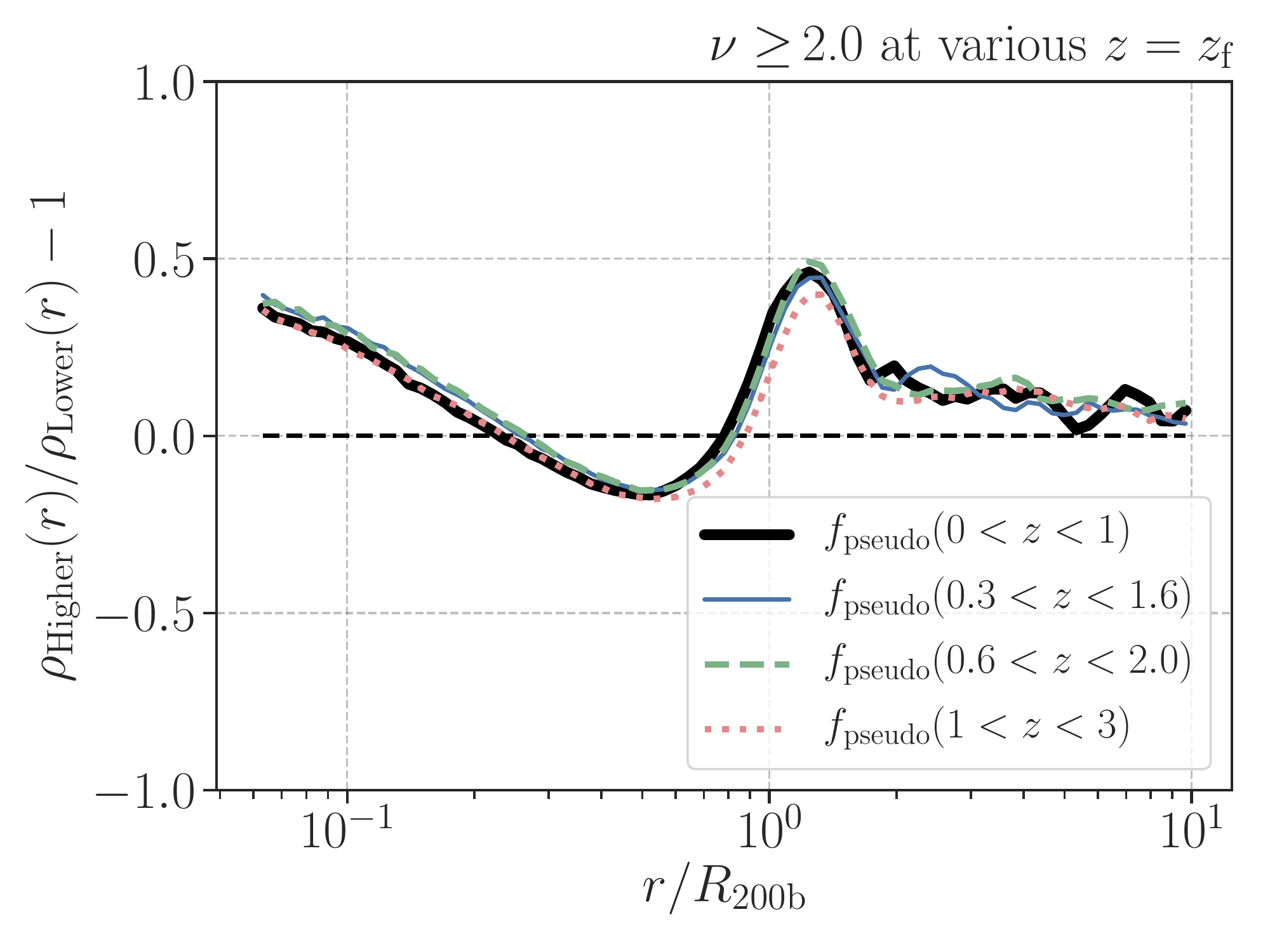}{0.4\textwidth}{}}
\caption{Similar to the bottom middle panel of Figure~\ref{fig:stack_prof}, but we here include the results at different redshifts. In this figure, we consider the halo sample with $\delta_{c}/\sigma(M_{\rm 200b}, z) \ge2$ at $z=0$, $0.3$, $0.6$, and $1$. For a given sample at $z=z_{\rm f}$, 
we measure the stacked density profiles of the halo sample with different selections for the amount of pseudo evolution since 
$z=z_{\rm i}$, denoted as $f_{\rm pseudo}(z_{\rm f}<z<z_{\rm i})$. 
The different lines in the figure show the fractional difference in the
stacked density profiles between two subsamples 
with $33\%$ higher and lower $f_{\rm pseudo}(z_{\rm f}<z<z_{\rm i})$ 
at various $z_{\rm f}$. \label{fig:stacked_prof_diffz}}
\end{figure}

Figure~\ref{fig:stacked_prof_diffz} summarizes 
the correlation between $f_{\rm pseudo}$ and the stacked density profiles 
at different redshifts \ms{when the minimum halo mass for each redshift is set at $\nu\ge2$}. 
\ms{We find the amount of pseudo evolution
and the stacked density profiles are correlated with each other in a similar manner 
even if the halo samples at $z>0$ are taken into consideration.}
We also confirm that the trend observed in Figure~\ref{fig:stacked_prof_diffz}
still remains if we vary the range in the peak height to be 
$\nu \ge 1.5$ or $\ge 2.5$.
\ms{Therefore, the relationship between the stacked density profile of the cluster-sized halos
and the amount of pseudo evolution is nearly universal across all masses and redshifts as long as
the halos are selected with respect to the peak height $\nu$.
%the stacked density profile of galaxy clusters
%depends on the amount of pseudo evolution 
%even at different redshifts 
%$0 \simlt z_{\rm f}\simlt1$. The dependence of $f_{\rm pseudo}$ is universal
%across masses and redshifts as long as the halos are selected with respect to the peak height $\delta_{c}/\sigma(M_{\rm 200b}, z)$.
}

\ms{In Appendix~\ref{app:frac_pseuod_vs_c_vs_b_diff_z}, we extensively study the correlation between the amount of pseudo evolution,
the linear bias, and the halo concentration at different redshifts and masses.
We find a similar trend to that shown in Figure~\ref{fig:c200b_bias} for various redshift and mass bins.}

%Figure~\ref{fig:c200b_bias_diff_z_m} shows the impact of amount of pseudo evolution on 
%the linear halo bias $b_{\rm L}$ and the concentration $R_{\rm 200b}/r_{s}$ at different redshifts and masses.
%We consider three bins in peak height parameter $\nu=\delta_{c}/\sigma(M_{\rm 200b},z)$ 
%at $z=0.3$, $0.6$ and 1.
%In Figure~\ref{fig:c200b_bias_diff_z_m}, we set the range of $\nu$ 
%to be $1.5\le\nu<2$, $2\le\nu<2.5$ and $\nu\ge2.5$ from left to right.
%For a given range of $\nu$ at redshift $z_{\rm f}$, 
%we divide the halo sample into three by the amount of pseudo evolution since $z=z_{\rm i} (=1+z_{\rm f})$.
%We then measure the linear halo bias $b_{\rm L}$ and the concentration $R_{\rm 200b}/r_{s}$
%of the subsamples as in Section~\ref{subsec:bias_and_c200}. 
%The upper (lower) triangles in each panel of Figure~\ref{fig:c200b_bias_diff_z_m}
%represent the fitting result of $b_{\rm L}$ and $R_{\rm 200b}/r_{s}$ for 
%the subsamples with $33\%$ higher (lower) value of the amount of pseudo evolution at different redshifts.
%The figure shows that the secondary bias by the amount of pseudo evolution can be more prominent 
%for the halos with smaller $\nu$ and all the subsamples with higher $f_{\rm pseudo}$ in this study
%tend to have larger concentration and reside in the environment with larger linear bias than the full sample.

\section{Discussion and Conclusion} \label{sec:concl}

In this study, we studied the mass accretion around the virial region 
of a sample of galaxy clusters with $N$-body simulations in detail.
Using a time series of the $N$-body simulation, we measured 
the redshift evolution of the spherical over-density (SO) mass of 
a galaxy cluster, along its mass accretion history.
Moreover, we evaluated the evolution of SO masses due to the change in a reference density, 
i.e. the pseudo evolution, on an individual basis at a galaxy-cluster scale.
The amount of pseudo evolution is expected to be important for a complete understanding of the mass evolution around galaxy clusters over a cosmic age.
Our findings are summarized as follows:

\begin{enumerate}

\item[(i)] For clusters with virial masses of $10^{14}\, h^{-1}M_{\odot}$ at $z=0$, 
\ms{$52\pm19\%$} the difference in the SO mass between $z=0$ and $1$ is 
accounted for by the pseudo evolution. 
The amount of pseudo evolution in a virial SO mass between $z=0$ and $1$, 
denoted as $f_{\rm pseudo}(0<z<1)$, generally decreases as $M_{\rm 200b}$ increases, 
where $M_{\rm 200b}$ is the SO mass with respect to 200 times the mean matter density at $z=0$.
The variance around the mean relationship between $f_{\rm pseudo}(0<z<1)$ and $M_{\rm 200b}$
is also found to decrease in more massive halos.

\item[(ii)] The amount of pseudo evolution between $z=0$ and $1$ can be 
correlated with a common estimation of halo age, the half-mass scale factor $a_{1/2}$, 
and the concentration of the inner density profile. In addition, we found a correlation 
between $f_{\rm pseudo}(0<z<1)$ and the tidal shear in the outskirts of galaxy clusters.
Our numerical simulation shows that halos with higher $f_{\rm pseudo}(0<z<1)$
tend to be older, more concentrated, and reside in environments 
with poorly developed filamentary structures.

\item[(iii)] 
\ms{The stacked density profile of a sample of galaxy clusters
depends on the amount of pseudo evolution between $z=0$ and $1$.}
The profile inside a virial radius of $z=0$ 
is more concentrated if one preferentially chooses galaxy clusters with higher $f_{\rm pseudo}(0<z<1)$.
Interestingly, we found that subsamples with higher $f_{\rm pseudo}(0<z<1)$ can show 
a larger amplitude in the stacked density profile beyond the virial radius.
This trend is a counter-example of the known secondary bias by halo concentration \citep[e.g.][]{2006ApJ...652...71W}
and is caused by the correlation between $f_{\rm pseudo}(0<z<1)$ and the tidal shear around clusters.
We estimated that the linear halo bias can be \ms{changed} at the $\sim6.25\%$ level due to 
the amount of pseudo evolution that has occurred in galaxy clusters with 
$M_{\rm 200b}=3.3\times10^{14}\, h^{-1}M_{\odot}$ at $z=0$.

\item[(iv)] We showed that the stacked density profile of galaxy clusters at $z>0$ is also 
\ms{correlated with} the amount of pseudo evolution within a finite redshift range.
For galaxy clusters at the four different redshifts $z=0$, $0.3$, $0.6$ and $1$, 
the impact of the amount of pseudo evolution on the stacked density profiles is found to 
be similar if the clusters are selected based on the peak height parameter $\nu = \delta_{c}/\sigma(M_{\rm 200b}, z)$, where $\delta_{c}=1.686$ and 
$\sigma$ is the top-hat mass variance at redshift $z$.
We also found the secondary bias due to the amount of pseudo evolution is more prominent 
in lower-$\nu$ halos than higher-$\nu$ halos.

\end{enumerate}

%Our results imply that the outer density profile around galaxy clusters would 
%depend on the presence of rich filamentary structures and tidal anisotropy of density environment.
Although our results concerning the amount of pseudo evolution are not expected to be significantly affected by complex 
baryonic feedback processes, 
the halo concentration of galaxy clusters would depend on the detailed astrophysics \citep[e.g.][]{2010MNRAS.405.2161D, 2012MNRAS.424.1244F, 2015MNRAS.451.1247S, 2018MNRAS.477.2804S}. 
\ms{In addition, it is still uncertain if baryons experience pseudo-evolution after accretion onto cluster-sized halos \citep[but see][for the recent study]{2015ApJ...808...40W}.}
Therefore, more thorough study with hydrodynamical simulations is required for the accurate modeling of halo concentration as a function of $f_{\rm pseudo}$.

Furthermore, the selection function of galaxy clusters at multiple wavelengths could be related to 
the amount of pseudo evolution of the SO masses. At least, the selection of galaxy clusters by weak lensing analysis will be subject to the amount of pseudo evolution, because 
the selection is based on the projected mass density around individual clusters \citep[e.g.][]{2004MNRAS.350..893H, 2012MNRAS.425.2287H, 2015MNRAS.453.3043S, 2018PASJ...70S..27M}.
We expect that clusters selected by lensing may preferentially have higher amounts of pseudo evolution,
because such clusters can have larger linear bias and concentration, and therefore the associated lensing signal of such structures can be boosted.
It would also be worth exploring whether the amount of pseudo evolution can be correlated 
with any other halo properties such as shape, the presence of substructures, spin, and anisotropy of the velocity.
These possible correlations would be important to understand the secondary bias on the scale of galaxy-clusters 
\citep[e.g.][]{2007MNRAS.377L...5G, 2010ApJ...708..469F, 2018MNRAS.474.5143M} and we leave these questions for future study.

% cosmological implication? how large biases in parameter estimation will be induced by ignoring f_pseudo

%\clearpage


\acknowledgments
The author thanks the anonymous referee for reading the paper carefully and providing thoughtful comments, many of which have resulted in changes to the revised version of the manuscript.
This work is in part supported by MEXT KAKENHI Grant Number (18H04358, 19K14767).
Numerical computations were in part carried out on Cray XC50 at Center 
for Computational Astrophysics, National Astronomical Observatory of Japan.



\appendix
\section{Systematic uncertainties in the estimate of amounts of pseudo evolution}\label{app:sys_err_f_pseudo}

\ms{
We here summarize the investigation of possible systematic effects on measuring the amount 
of pseudo evolution using a time series of $N$-body simulations.
}

\begin{figure*}[t!]
%\gridline{\fig{f_pseudo_err_estimate.pdf}{1.\textwidth}{}}
\gridline{\fig{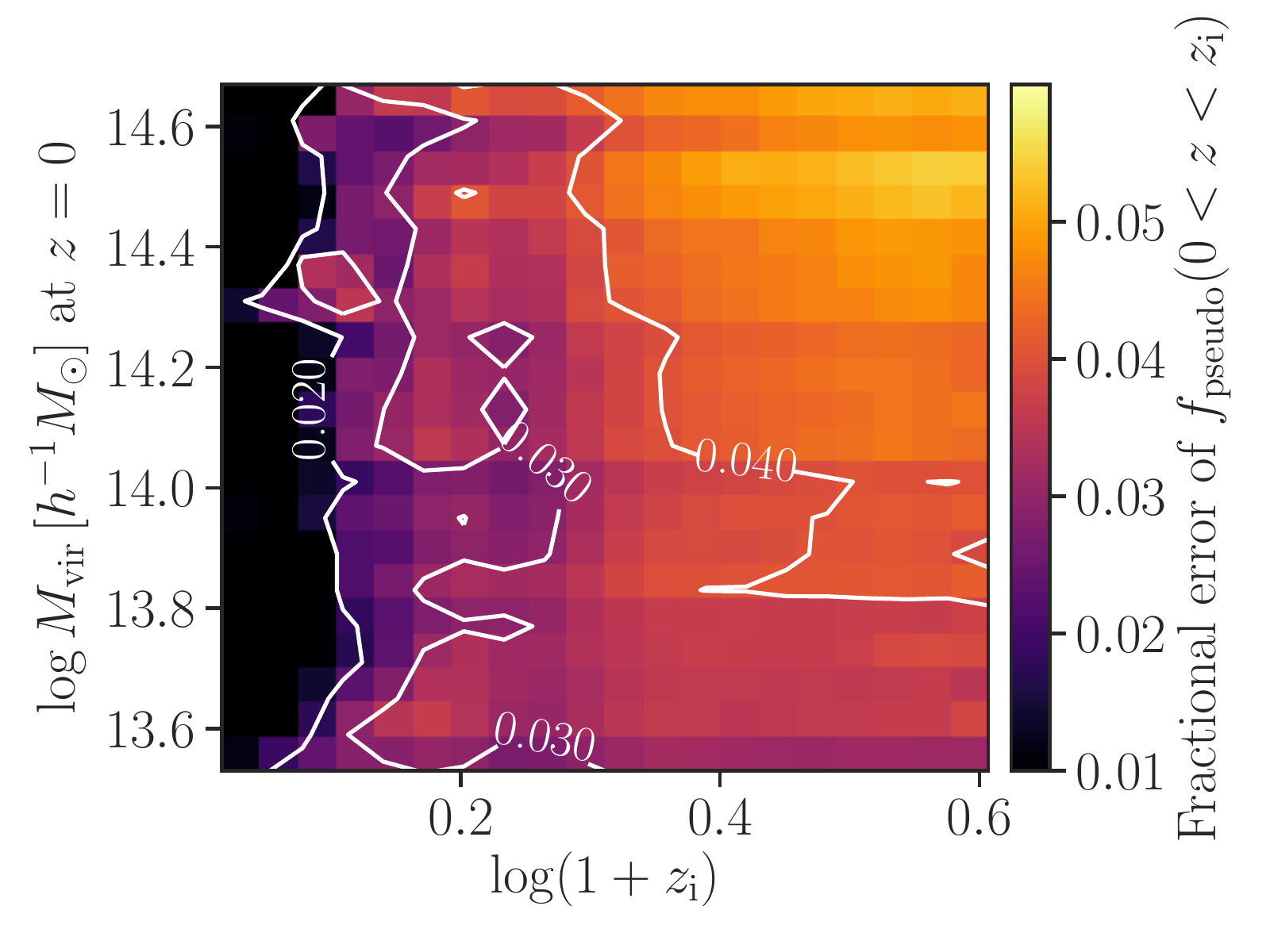}{0.5\textwidth}{}
          \fig{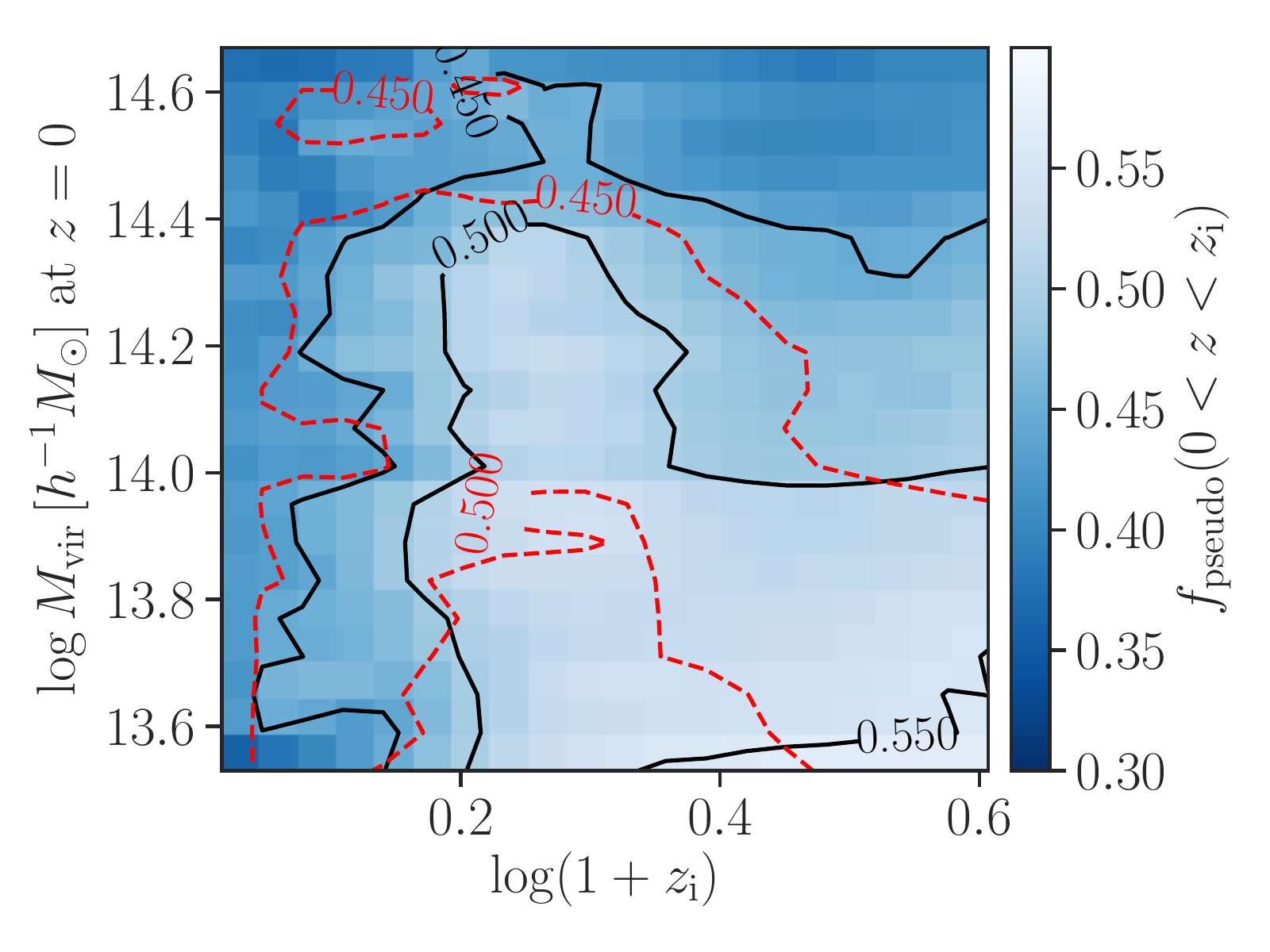}{0.5\textwidth}{}
          }
\caption{
Systematic errors on the estimation of the amount of pseudo evolution $f_{\rm pseudo}$. 
The left panel summarizes the systematic effect arising from the time evolution of the mass density profiles
between two snapshots in our numerical simulations, while the right represents the impact of the contributions from the outer mass profile beyond the virial radius on the estimate of $f_{\rm pseudo}$.
In the left, we show the fractional difference between the lower and upper limits of $f_{\rm pseudo}$,
assuming that the density profiles grow monotonically at all radii. In the right panel, the colored map and the solid contours show $f_{\rm pseudo}$ as a function of the initial redshift $z_{i}$ and the virial SO mass at $z=0$, while the red dashed contours represent the estimates when we use the NFW profiles 
for individually simulated halos.
\label{fig:f_pseudo_err}}
\end{figure*}

\ms{
As in Section~\ref{subsec:stats}, we estimate the amount of pseudo evolution of the virial SO mass
assuming that the mass density profiles can be almost static between two snapshots in our simulations.
Under this static-profile approximation, we can write the pseudo-evolved masses from $z=z_{1}$
to $z_{2} > z_{1}$ as Eq.~(\ref{eq:delta_M_pseudo}). Similarly, we can estimate the pseudo-evolved masses
in descending order of redshift as
\beqa
\Delta M^{\prime}_{\rm pseudo}(z_2) \equiv M(<R_{{\rm vir}, 1}, z_2) - M(<R_{{\rm vir}, 2}, z_2), \label{eq:delta_M_pseudo_prime}
\eeqa
As discussed in \citet{2013ApJ...766...25D}, assuming the mass density profiles evolve monotonically
at all radii between two redshifts, Eqs.~(\ref{eq:delta_M_pseudo}) and (\ref{eq:delta_M_pseudo_prime})
can provide the upper and lower limits of the pseudo-evolved masses, respectively.
Hence, we introduce the following quantity to assess the systematic error 
in the estimate of $f_{\rm pseudo}$,
\beqa
\Delta f_{\rm pseudo}(0<z<z_{\rm i}) &\equiv& 1 - \frac{\sum_{0<z<z_{\rm i}}M^{\prime}_{\rm pseudo}(z)}{\sum_{0<z<z_{\rm i}}M_{\rm pseudo}(z)}.
\eeqa
The left panel of Figure~\ref{fig:f_pseudo_err} shows $\Delta f_{\rm pseudo}(0<z<z_{\rm i})$
as a function of the initial redshift $z_{\rm i}$ and the virial SO mass at $z=0$.
We find that the time evolution in the mass density profiles between two snapshots can 
a $3-4\%$ effect our estimates of $f_{\rm pseudo}$ at $z_{\rm i}=1$.
}

\ms{
Apart from the time evolution, we also study the impact of the outer mass profiles on the estimate of $f_{\rm pseudo}$. For this purpose, we use another estimation of the pseudo-evolved masses based on
the Navarro-Frenk-White (NFW) profiles (Eq.~[\ref{eq:NFW}]) for individual halos, 
\beqa
\Delta M_{\rm pseudo, NFW}(z_1) \equiv M_{\rm NFW}(<R_{{\rm vir}, 1}, z_1) - M_{\rm NFW}(<R_{{\rm vir}, 2}, z_1), \label{eq:delta_M_pseudo_NFW} 
\eeqa
where $M_{\rm NFW}(<R, z)$ represents the enclosed mass profile given by the NFW profile.
To compute Eq.~(\ref{eq:delta_M_pseudo_NFW}), we use the scale radii ($r_s$) and the virial SO masses ($M_{\rm vir}$) obtained from the {\tt ROCKSTAR} algorithm on a halo-by-halo basis.
In the right panel of Figure~\ref{fig:f_pseudo_err},
the color map and the solid contours represent our fiducial estimate of $f_{\rm pseudo}$ (Eq.~[\ref{eq:f_pseudo}]) as a function of $z_{i}$ and the present-day virial SO mass,
while the red dashed counters are the counterparts based on Eq.~(\ref{eq:delta_M_pseudo_NFW}).
The NFW-profile-based estimates are broadly consistent with the results in \citet{2015ApJ...810...36M}.
We found that the outer mass profile directly extracted from the simulations 
can affect the estimate of $f_{\rm pseudo}$ at $z_{\rm i}=1$ by $\sim10\%$.
}

\section{The pseudo evolution of SO mass with respect to 200 times mean density}\label{app:f_pseudo_m200b}

\begin{figure*}[t!]
%\plotone{hist_frac_pseudo_z0.pdf}
\gridline{\fig{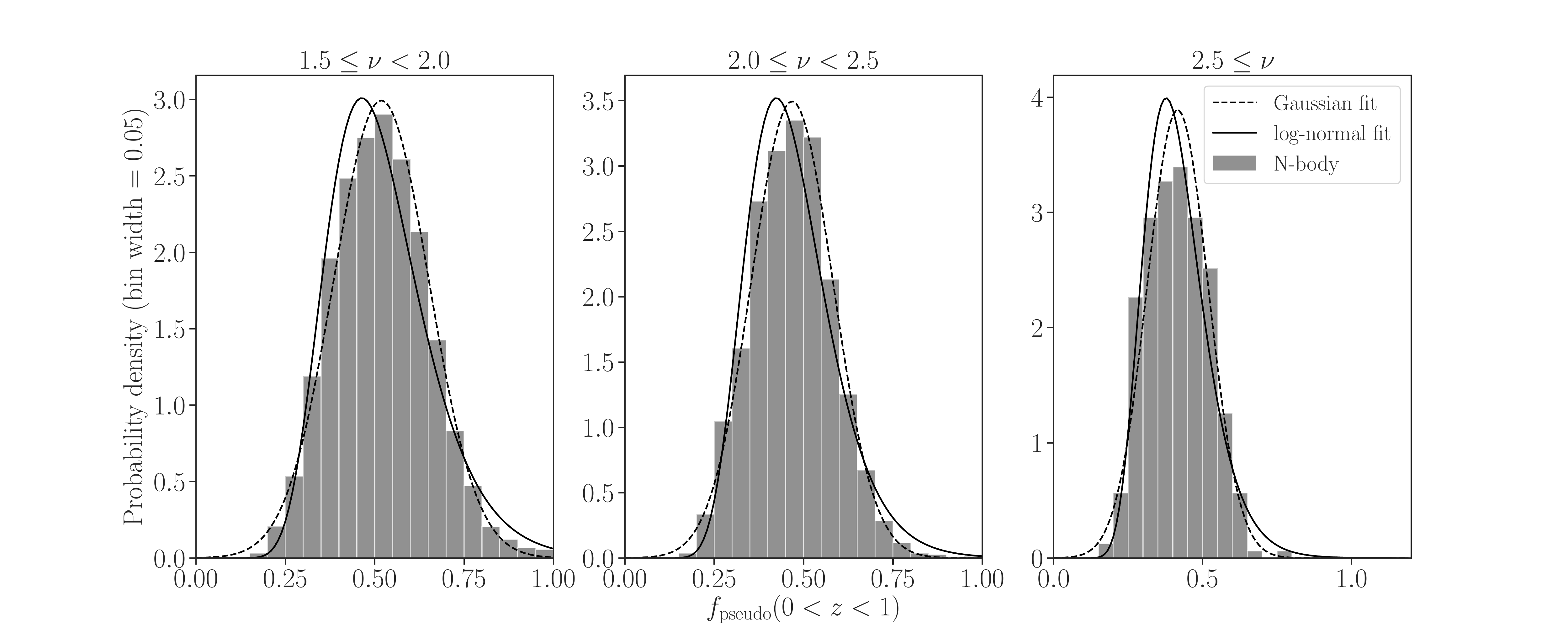}{1.\textwidth}{}}
\caption{Similar to Figure~\ref{fig:hist_frac_pseudo_z0}, but we define the fraction of pseudo mass evolution
by using the SO mass with respect to 200 times mean mass density. \label{fig:hist_frac_pseudo_z0_m200b}}
\end{figure*}

\ms{
In this Appendix, we examine the amount of pseudo evolution of the SO mass with respect to 200 times mean matter density $M_{\rm 200b}$. We adopt a similar estimation to that used in Eq.~(\ref{eq:f_pseudo}), but replace the virial masses and radii with $M_{\rm 200b}$ and $R_{\rm 200b}$, respectively. 
As in the main text, we consider halo samples with three different bins of peak height $\nu$, $1.5\le\nu<2$, $2\le\nu<2.5$, and $2.5\le\nu$.
}

\ms{
Figure~\ref{fig:hist_frac_pseudo_z0_m200b} shows the histogram of $f_{\rm pseudo}(0<z<1)$ when 
using the SO mass of $M_{\rm 200b}$ for our three subsamples at $z=0$. 
In each panel, the gray boxes represent the simulation results, while the solid and dashed lines
are the Gaussian fit and log-normal fits, respectively. 
Notably, the distribution of $f_{\rm pseudo}(0<z<1)$ becomes narrower when the SO mass definition of $M_{\rm 200b}$ is used than for the $M_{\rm vir}$-counterparts, leading to simple Gaussian fits that can explain the simulation results.
%The fitting results are provided in Table~\ref{tab:table_f_mean_std_m200b}.
We find the relationship between the averaged $f_{\rm pseudo}(0<z<1)$ for $M_{\rm 200b}$ can be approximated by
\beqa
\langle f_{\rm pseudo}(0<z<1) \rangle &=& 
(-0.0186\pm0.0009) M_{\rm 200b, 14} + (0.5311\pm0.0016), \label{eq:f_mean_m200b}
\eeqa
where 
$M_{\rm 200b, 14} =\left( M_{\rm 200b}/10^{14}\, h^{-1}M_{\odot} \right)$.
%and we find a clear decreasing trend of $f_{\rm pseudo}$ as a function of halo mass.
In addition, the variance around Eq.~(\ref{eq:f_mean_m200b}) is given by 
$(-0.0014\pm0.0001) M_{\rm 200b, 14} + (0.0185\pm0.0003)$.
%showing more massive galaxy cluster has a smaller scatter in $f_{\rm pseudo}$.
}

\section{The Connection of the amount of pseudo evolution to the linear bias and the concentration 
at various redshifts}\label{app:frac_pseuod_vs_c_vs_b_diff_z}

\begin{figure*}[t!]
\gridline{\fig{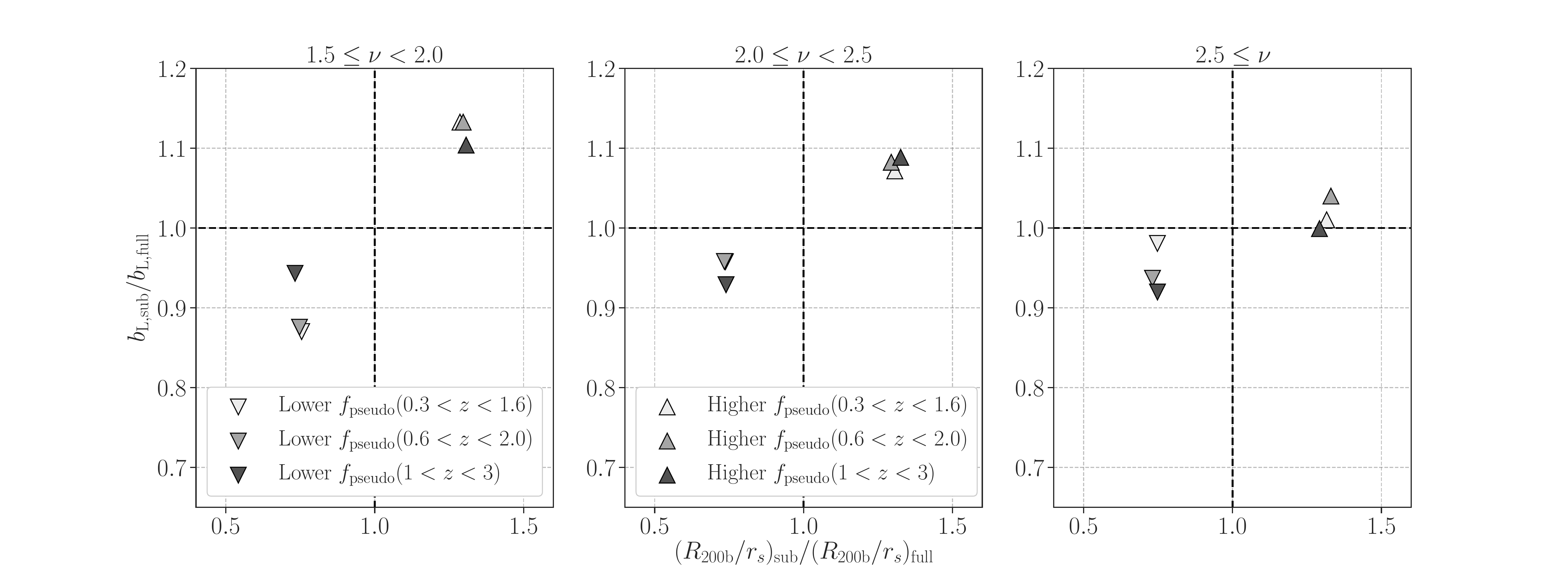}{1.\textwidth}{}}
\caption{Similar to Figure~\ref{fig:c200b_bias}, 
but here we show the results at different redshifts and halo masses.
In this figure, we consider three different ranges of peak height $\nu=\delta_{c}/\sigma(M_{\rm 200b},z)$.
From left to right, the corresponding $\nu$ range is given by $1.5\le\nu<2$, 
$2\le\nu<2.5$ and $\nu\ge2.5$, respectively.
For a given $\nu$ range at $z=z_{\rm f}$, 
we divide the halo sample by the amount of pseudo evolution between $z=z_{\rm f}$ and $z_{\rm i}$ 
(denoted as $f_{\rm pseudo}(z_{\rm f}<z<z_{\rm i})$)
and derive the linear halo bias $b_{\rm L}$ and the halo concentration $R_{\rm 200b}/r_s$.
Each panel summarizes the fitted result of $b_{\rm L}$ and $R_{\rm 200b}/r_s$ for 
the subsample with $33\%$ higher and lower $f_{\rm pseudo}$. Note that the bias and the halo concentration in this figure are normalized by the counterparts for the full sample.
\label{fig:c200b_bias_diff_z_m}}
\end{figure*}

\ms{
Figure~\ref{fig:c200b_bias_diff_z_m} shows the impact of the amount of pseudo evolution on 
the linear halo bias $b_{\rm L}$ and the concentration $R_{\rm 200b}/r_{s}$ at different redshifts and masses.
We consider three bins of peak height parameter $\nu=\delta_{c}/\sigma(M_{\rm 200b},z)$ 
at $z=0.3$, $0.6$, and $1$.
In Figure~\ref{fig:c200b_bias_diff_z_m}, we set the range of $\nu$ 
to be $1.5\le\nu<2$, $2\le\nu<2.5$, and $\nu\ge2.5$ from left to right.
For a given range of $\nu$ at redshift $z_{\rm f}$, 
we divide the halo sample into three by the amount of pseudo evolution as $z=z_{\rm i}$.
The redshift $z_{\rm i}$ is set to ensure the cosmic time interval between $z_{\rm i}$ and $z_{\rm f}$
is equal to double that of the dynamical time at redshift $z_{\rm f}$.
We then measure the linear halo bias $b_{\rm L}$ and the concentration $R_{\rm 200b}/r_{s}$
of the subsamples as in Section~\ref{subsec:bias_and_c200}. 
The upper (lower) triangles in each panel of Figure~\ref{fig:c200b_bias_diff_z_m}
represent the fitting results of $b_{\rm L}$, and $R_{\rm 200b}/r_{s}$ for 
the subsamples with $33\%$ higher (lower) values of the amount of pseudo evolution at different redshifts.
The figure shows that the secondary bias by the amount of pseudo evolution can be more prominent 
for the halos with smaller $\nu$ and all the subsamples with higher $f_{\rm pseudo}$ in this study 
tend to be of larger concentrations and reside in the environment with a greater linear bias than the full sample.
}

\bibliographystyle{aasjournal}
\bibliography{refs} % if your bibtex file is called example.bib



\end{document}